\documentclass[superscriptaddress,nobibnotes,amsmath,amssymb,notitlepage,twocolumn,pra,longbibliography]{revtex4-2}

\usepackage{titlesec}
\titleformat{\section}
  {\centering\normalfont\fontsize{12}{15}\bfseries}{\thesection}{1em}{}
\makeatletter
\def\@hangfrom@section#1#2#3{\@hangfrom{#1#2}#3}
\def\@hangfroms@section#1#2{#1#2}
\makeatother

\usepackage{bm,braket}
\usepackage[toc,page]{appendix}
\usepackage{comment}
\usepackage{dcolumn}

\usepackage{amsfonts}

\usepackage{graphicx,color,hyperref}
\usepackage[caption=false]{subfig}
\hypersetup{colorlinks=true, linkcolor=blue, citecolor=blue, urlcolor=blue} 

\newcommand{\beq}[1]{\begin{equation}\label{#1}}
\newcommand{\eep}{\;.\end{equation}}
\newcommand{\eec}{\;,\end{equation}}
\newcommand{\eeq}{\end{equation}}

\newcommand*\dd{\mathop{}\!\mathrm{d}} 

\makeatletter
\newcommand{\fmarki}{*}
\newcommand{\fmarkii}{\ensuremath{\dagger}}
\newcommand{\fmarkiii}{\ensuremath{\ddagger}}
\newcommand{\fmarkiv}{\ensuremath{\mathsection}}
\newcommand{\fmarkv}{\ensuremath{\mathparagraph}}
\newcommand{\fmarkvi}{\ensuremath{\|}}
\newcommand{\fmarkvii}{**}
\newcommand{\fmarkviii}{\ensuremath{\dagger\dagger}}
\newcommand{\fmarkix}{\ensuremath{\ddagger\ddagger}}
                
\def\@fnsymbol#1{{\ifcase#1\or \fmarki\or \fmarkii\or \fmarkiii\or \fmarkiv\or \fmarkv\or \fmarkvi\or \fmarkvii\or \fmarkviii\or \fmarkix \else\@ctrerr\fi}}
\makeatother

\renewcommand{\fmarki}{$*$}
\renewcommand{\fmarkii}{$*$}
\renewcommand{\fmarkiii}{$*$}
\renewcommand{\fmarkiv}{$*$}
\renewcommand{\fmarkv}{$*$}
\renewcommand{\fmarkix}{$*$}

\DeclareMathAlphabet{\mathcal}{OMS}{cmsy}{m}{n} 

\renewcommand{\vec}[1]{{\bf #1}}

\newcommand{\kv}{\vec{k}}

\usepackage{amsmath}
\usepackage{amssymb}
\usepackage{xcolor}
\usepackage{bbm}
\usepackage{physics}
\usepackage{float}
\usepackage{graphicx}
\usepackage{dcolumn} 
\usepackage{bm} 
\usepackage{siunitx}
\usepackage{enumitem}  

\makeatletter
\renewcommand*{\fnum@figure}{{\normalfont\bfseries \figurename~\thefigure}}
\makeatother

\allowdisplaybreaks

\definecolor{orange}{rgb}{1,0.5,0}

\graphicspath{{./img/}}

\DeclareMathAlphabet{\mathcal}{OMS}{cmsy}{m}{n} 

\newcommand{\ii}{\mathrm{i}}

\makeatletter
\newcommand{\specificthanks}[1]{\@fnsymbol{#1}}
\makeatother

\begin{document}

\preprint{APS/123-QED}

\title{Excitonic topology and quantum geometry in organic semiconductors}

\author{Wojciech J. Jankowski}
\email{wjj25@cam.ac.uk}
\thanks{}
\affiliation{TCM Group, Cavendish Laboratory, Department of Physics, J J Thomson Avenue, Cambridge CB3 0HE, United Kingdom}

\author{Joshua J.P. Thompson}
\thanks{}
\affiliation{Department of Materials Science and Metallurgy, University of Cambridge,
27 Charles Babbage Road, Cambridge CB3 0FS, United Kingdom}

\author{Bartomeu Monserrat}
\thanks{}
\affiliation{TCM Group, Cavendish Laboratory, Department of Physics, J J Thomson Avenue, Cambridge CB3 0HE, United Kingdom}
\affiliation{Department of Materials Science and Metallurgy, University of Cambridge,
27 Charles Babbage Road, Cambridge CB3 0FS, United Kingdom}

\author{Robert-Jan Slager}
\email{rjs269@cam.ac.uk}
\affiliation{TCM Group, Cavendish Laboratory, Department of Physics, J J Thomson Avenue, Cambridge CB3 0HE, United Kingdom}
\affiliation{Department of Physics and Astronomy, University of Manchester, Oxford Road, Manchester M13 9PL, United Kingdom}

\date{\today}

\begin{abstract}
\section*{Abstract}
Excitons drive the optoelectronic properties of organic semiconductors which underpin devices including solar cells and light-emitting diodes. Here we show that excitons can exhibit topologically non-trivial states protected by inversion symmetry and identify a family of organic semiconductors realising the predicted excitonic topological phases. We also demonstrate that the topological phase can be controlled through experimentally realisable strains and chemical functionalisation of the material. Appealing to quantum Riemannian geometry, we predict that topologically non-trivial excitons have a lower bound on their centre-of-mass spatial spread, which can significantly exceed the size of a unit cell. Furthermore, we show that the dielectric environment allows control over the excitonic quantum geometry. The discovery of excitonic topology and excitonic Riemannian geometry in organic materials brings together two mature fields and suggests many new possibilities for a range of future optoelectronic applications.  
\end{abstract} 

\maketitle

\section*{Introduction}
Topology represents a versatile tool that drives the study of diverse condensed matter phenomena. Topological invariants arise due to phases acquired by wave functions when adiabatically transported around the Brillouin zone (BZ), and provide a classification of states of matter that has radically transformed our understanding of inorganic materials over the past few decades\,\cite{Rmp1,Rmp2}. This work has culminated in a rather complete understanding of free fermionic band structures, which can be characterised through the gluing of symmetry eigenvalues, or irreducible representations of eigenstates, between high symmetry points in the momentum space Brillouin zone\,\cite{Clas1, Clas2, Clas3, FuKane3D}. Assessing whether this admits an exponentially localised Wannier representation in real space leads to their topological characterisation\,\cite{Clas4, Clas5}. Current efforts are moving from free fermion systems towards the question of how these ideas can drive new understanding in interacting systems. An example of an interacting system is that of electron-hole bound states, or excitons. 

Excitons dominate the optoelectronic properties of a wide range of materials, particularly low-dimensional materials and materials composed of organic molecules. Organic semiconductors, in particular, possess excellent light-harvesting properties driven by the formation of excitons\,\cite{kasha1965exciton, mikhnenko2015exciton, cocchi2018polarized, thompson2023singlet}, while being chemically versatile\,\cite{davies2021radically, cocchi2018polarized}, cheap, and environmentally friendly to fabricate\,\cite{allard2008organic, darling2013case}. These features make organic semiconductors one of the most promising material platforms in which to realise optoelectronic devices, from photovoltaics\,\cite{congreve2013external, sun2022recent} and light-emitting diodes\,\cite{kim2000electroluminescence,cho2024suppression} to biosensors\,\cite{jacoutot2022infrared}. The electron-hole distance and the centre-of-mass location of excitons can differ greatly between materials, and delocalised excitons with high mobilities are particularly promising for applications\,\cite{sneyd2022new}. 

\begin{figure} [!t]
    \centering
    \includegraphics[width=\linewidth]{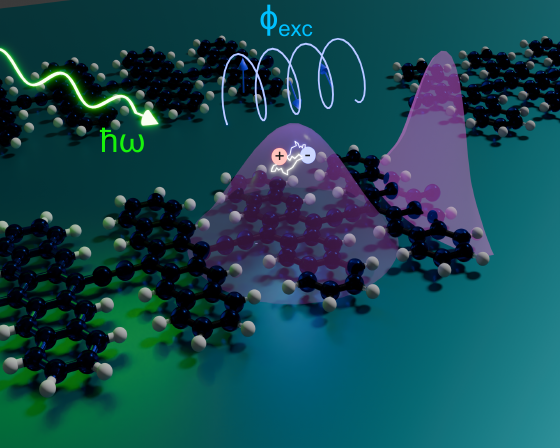}
    \caption{{\bf Topological excitons in organic polymers.} Formation of topological excitons (purple) from electron (blue) and hole (red) pairs in organic materials, induced by photoexcitation with photons of energy $\hbar \omega$ (green). Topological excitons exhibit a winding of the excitonic Bloch vectors $\ket{u^{\text{exc}}_{\textbf{Q}}}$ as the momentum of the excitons is changed, which is captured by the excitonic Berry phases $\phi_{\text{exc}}$ and associated inversion symmetry-protected topological $\mathbb{Z}_2$ invariant $P_{\text{exc}}$.}
    \label{fig:fig1}
\end{figure}

In this work, we establish a link between the rich exciton phenomenology in organic semiconductors with topological insights of such phases, developing a promising platform in which these ideas could culminate in experimentally viable topological phases and phenomena. In particular, we fully characterise the inversion-symmetry-protected excitonic topology in one-dimensional systems in terms of the excitonic Berry phase and a $\mathbb{Z}_2$ excitonic topological invariant. Additionally, we derive a lower bound on the spread of the exciton centre-of-mass wavefunction arising from a non-trivial excitonic quantum metric induced by the excitonic topology. We~derive these results using model Hamiltonians and, importantly, we also predict their manifestation in the polyacene family of one-dimensional organic semiconductors. We identify material realisations of predicted topological excitonic phases, which can be manipulated both chemically through the lateral size of the polyacene chains and externally through strain and through changes in the dielectric environment. Our findings of exotic excitonic states in organic materials bring together organic chemistry, semiconductor physics, and topological condensed matter, as illustrated in Fig.~\ref{fig:fig1}.

\section*{Results}

\subsection*{Excitonic topology and Berry phases}
The topological character of excitons is fully encoded in the excitonic wavefunction $\ket{\psi^{\text{exc}}_\textbf{Q}}$. In a periodic crystal, the excitonic wavefunction\,\cite{Yao2008, Brey2021, Kwan2021, MaiselLiceran2023, Xie2024} can be decomposed in terms of (i) electronic single-particle Bloch states ${\ket{\psi^{\text{e}}_\textbf{k}} = e^{i\kv \cdot \vec{r}_{\text{e}}} \ket{u^{\text{e}}_\textbf{k}}}$, (ii) hole single-particle Bloch states ${\ket{\psi^{\text{h}}_\textbf{k}} = e^{i\kv \cdot \vec{r}_{\text{h}}} \ket{u^{\text{h}}_\textbf{k}}}$, and (iii) an interaction-dependent envelope function $\psi_{\textbf{Q}}(\kv)$; leading to a coherent superposition of electron-hole pairs: 
\beq{eq::decomposition}
    \ket{\psi^{\text{exc}}_\textbf{Q}} = \sum_\kv \psi_{\vec{Q}}(\kv) e^{i\kv \cdot (\vec{r}_{\text{e}} - \vec{r}_{\text{h}})} \ket{u^{\text{e}}_{\mathbf{k+Q}/2}} \ket{u^{\text{h}}_{\mathbf{-k+Q}/2}}.
\eeq
We also introduce $\ket{u^{\text{exc}}_\textbf{Q}}$ as the cell-periodic part of the excitonic Bloch state $\ket{\psi^{\text{exc}}_{\textbf{Q}}} = e^{i\vec{Q}\cdot\vec{R}} \ket{u^{\text{exc}}_\textbf{Q}}$, where $\vec{R} = (\vec{r}_e  + \vec{r}_h)/2$ is the centre-of-mass position of the exciton. Notably, the relative position $\vec{r} = \vec{r}_{\text{e}} - \vec{r}_{\text{h}}$ of the electron and hole  enters the excitonic state via phase factors weighted by the envelope amplitudes $\psi_{\textbf{Q}}(\textbf{k})$. The solutions of the single-particle electronic problem for electrons and holes, and of a two-body equation for the envelope part, fully determine the excitonic wavefunction. The excitonic topology can then be studied from the cell-periodic part $\ket{u^{\text{exc}}_\textbf{Q}}$, on eliminating the phase factor with the centre-of-mass momentum $\textbf{Q}$ coupling to the centre-of-mass position $\vec{R}$. 

Focusing on one spatial dimension and on the associated one-dimensional momentum space with exciton momenta $\vec{Q} = Q$, K-theory\,\cite{Kitaevtenfold,Clas3}, which provides classifications that are stable up to adding an arbitrary number of trivial bands to the system, dictates that in the presence of inversion symmetry we can obtain an excitonic $\mathbb{Z}_2$ invariant $P_{\text{exc}} = \phi_{\text{exc}}/\pi\,\text{mod}\,2$ from a Berry phase\,\cite{Zak1989}:
\beq{eq::excBerry} 
    \phi_{\text{exc}} = \int_{\text{BZ}} \dd Q~ A_{\text{exc}}(Q),
\eeq
where $A_{\text{exc}}(Q) = \ii\bra{{u^{\text{exc}}_Q}}\ket{\partial_Q u^{\text{exc}}_Q}$ is the excitonic Berry connection. 
Alternatively, the excitonic $\mathbb{Z}_2$ invariant can be rewritten as\,\cite{PhysRevB.83.245132}:
\beq{}
    P_{\text{exc}} = \frac{1}{\pi} \int_\text{hBZ} \dd Q~[ A_{\text{exc}}(Q) + A_{\text{exc}}(-Q)],
\eeq
with the integration performed over a half Brillouin zone (hBZ) because the inversion symmetry relates the excitonic bands at $Q$ and $-Q$. The $\mathbb{Z}_2$ nature of the invariant can be understood intuitively: at the inversion-symmetry-invariant momenta, $Q=0$ and $Q=\pi$, the excitonic eigenvectors $\ket{u^{\text{exc}}_Q}$ have the same relative phase in the trivial excitonic regime, whereas in the topological regime these have opposite inversion eigenvalues related by a $\pi$ phase. The invariant $P_{\text{exc}}$ precisely distinguishes two different topologies which are not transversable from one to another without closing a gap in the excitonic bands, similarly to the inversion-protected topology of electrons necessitating a gap closure to change the Berry phase, and for a topological phase transition to occur.

\subsection*{Modelling excitons in one dimension}
To explore excitonic topology in a one-dimensional setting, we consider the excitonic Wannier equation for the envelope function $\psi_{Q}(k)$:
\beq{}
    \sum_{k'} h_{k,k'}(Q) \psi_{Q}(k) = E(Q) \psi_{Q}(k),
\eeq
where $E(Q)$ is the exciton binding energy and $h_{k,k'}(Q) = (E^e_{k+Q/2} -  E^h_{k-Q/2} ) \delta_{k, k'} - W_{k,-k', Q}$ includes the single-particle energies of the electron $E^e_{k+Q/2}$ and hole $E^h_{k-Q/2}$ bands, and the electron-hole interaction $W_{k,-k', Q}$. 

For the single-particle electron and hole states, which we emphasise, only form a part of the model, we use the Su-Schrieffer-Heeger (SSH) model\,\cite{ssh1,ssh2}:
\beq{eq::elecSSH}
\begin{split}
    H = -t_{1} \sum_{j} c^{\dagger}_{B,j} c_{A,j} -t_{2} \sum_{j} c^{\dagger}_{{B+1},j} c_{A,j} + \text{h.c.},
\end{split}
\eeq
with $c^{\dagger}_{B,j}/c_{A,j}$ the creation/annihilation operators for the electrons at sublattices $A, B$, in unit cell $j$, and effective staggered hopping parameters $t_1$ and $t_2$. We define the origin of a unit cell such that $t_1$ is the intracell hopping and $t_2$ is the intercell hopping across the boundary of the unit cell.

\subsection*{Characterisation of topological excitons}
\begin{figure} [!t]
    \centering \includegraphics[width=0.8\linewidth]{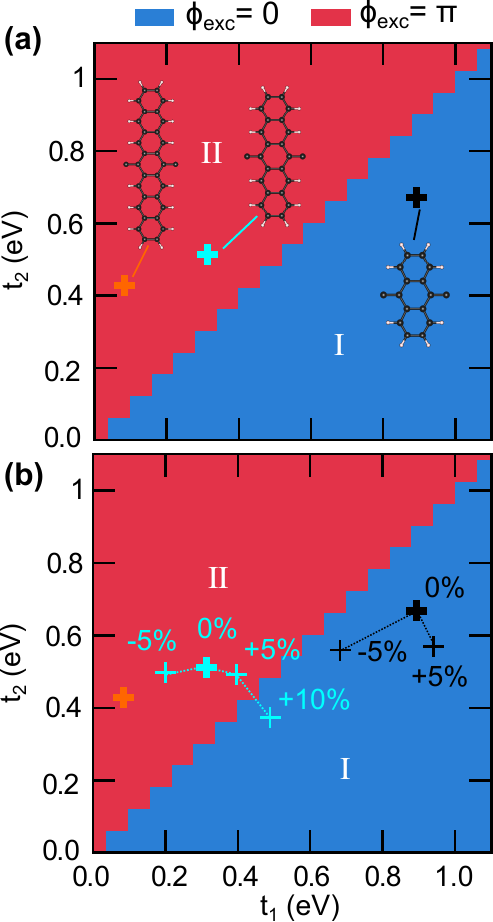}
    \caption{{\bf Topological excitonic phase diagrams.}
    (\textbf{a}) Phase diagram showing excitonic topological phases as a function of $t_1$ and $t_2$, with $L=0.8$\,nm and $\alpha_{\text{1D}} = 0.05$\,eV\AA, see Methods. The phase regimes are defined as I, II, as discussed in the main text. We also indicate the location of polyanthracene (black cross), polypentacene (cyan cross) and polyheptacene (orange cross) in the phase diagram. (\textbf{b}) Controllability of excitonic topology with strain. We explore the excitonic topological phase diagram of the lowest excitonic band on applying uniform strains within a range $\gamma = \pm 10 \%$ of the relative polymer length change. We find that the excitonic phases can evolve from region II to region I, demonstrating that the non-trivial excitonic topology can be trivialised with strain $\gamma < 10 \%$.}
    \label{fig:fig2}
\end{figure}

By numerically solving the Wannier equation, we construct the excitonic phase diagram of our fully interacting problem as a function of single-particle states characterised by the hopping parameters $t_1$ and $t_2$, as shown in Fig.\,\ref{fig:fig2}(a). The phase diagram exhibits two different regimes, one corresponding to trivial excitonic topology and the other corresponding to non-trivial excitonic topology.

In regime I, corresponding to $t_1>t_2$ in Fig.\,\ref{fig:fig2}, we have trivial electrons and holes yielding trivial excitons. This is a consequence of the absence of winding in the vectors $\ket{u^{\text{e}}_k}$ and $\ket{u^{\text{h}}_k}$ that leads to a trivial electronic Berry phase with $\phi_{\text{e}} = 0$. This regime is associated with both electron and excitonic Wannier centres localised at the centres of the unit cells, see Fig.\,\ref{fig:fig3}(a). 

In regime II, corresponding to $t_1<t_2$, we have topological electrons and holes, resulting in topological excitons, whose topological character is inherited from the underlying electron and hole topology. Explicitly, we have that the single-particle Berry phases $\phi_{\text{e}}$ for electrons, $\phi_{\text{h}}$ for holes, and the excitonic Berry phase $\phi_{\text{exc}}$, obey $ \phi_{\text{exc}} = \phi_{\text{e}} = \phi_{\text{h}} = \pi$. This regime is associated with both electron and exciton Wannier centres localised at the edges of the unit cells, see Fig.\,\ref{fig:fig3}(b).

\subsection*{Phenomenological understanding of the phase diagram}
\begin{figure} [!t]
    \centering
    \includegraphics[width=0.9\linewidth]{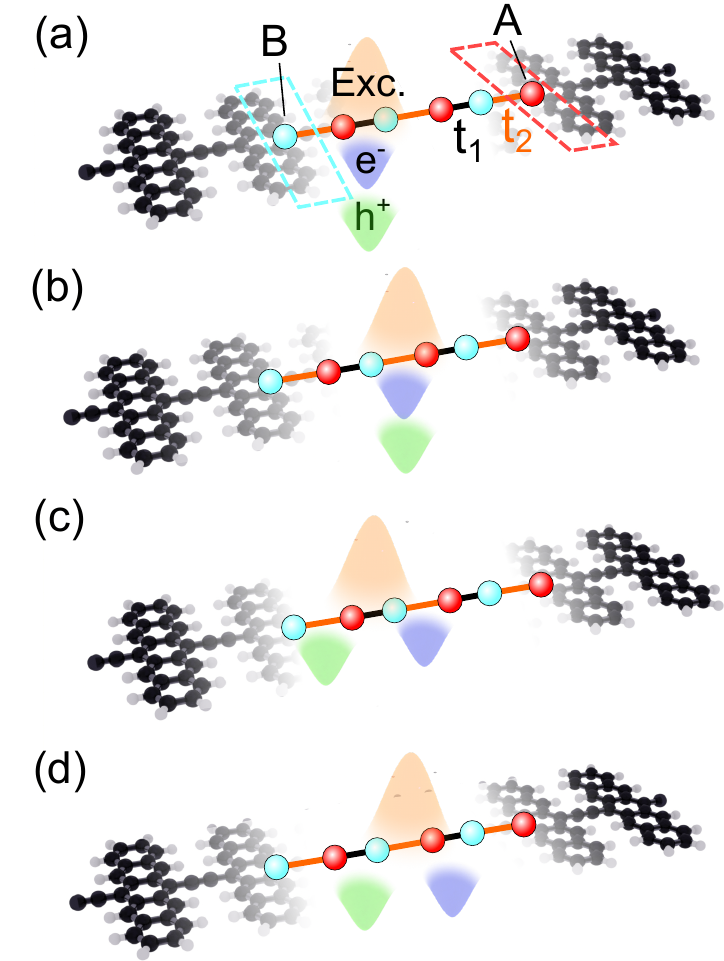}
    \caption{{\bf Topological excitons in real space.} Real space realisation of different combinations of Wannier centre shifts with respect to sublattice sites $A$, $B$ admitting electron hopping amplitudes $t_1, t_2$. The Wannier centre shifts correspond to the Berry phases of electrons  ($e^-$), holes ($h^+$), and excitons (\text{Exc.}) shown in blue, green, and orange, respectively. The interaction-driven configurations of the formation of topological and trivial excitons are shown schematically. 
    \textbf{(a)} Trivial electrons and holes forming trivial excitons ($P_{\text{exc}} = 0$) in few-ring polyacenes. \textbf{(b)} Topological electrons and holes forming topological excitons ($P_{\text{exc}} = 1$) in many-ring polyacenes. The other possibilities include:  \textbf{(c)} Topological electrons and holes forming trivial excitons ($P_{\text{exc}} = 0$). \textbf{(d)} Topological excitons ($P_{\text{exc}} = 1$) formed out of trivial electrons and holes, as  conjectured in Ref.~\cite{davenport2024interactioninduced}. We find that only cases \textbf{(a-b)} can be realised in the excitonic bands of polyacenes. 
    }
    \label{fig:fig3}
\end{figure}

To gain a deeper understanding of the excitonic topology and phase diagram presented in Figs.\,\ref{fig:fig2} and \ref{fig:fig3}, we now show that the electron-hole interaction can be phenomenologically captured through a dualisation picture using a SSH-Hubbard-like model with a hierarchy of electronic density-density interactions $U_i$, with ${i = 1,2,3, \ldots}$ labelling interactions between $n$th order left neighbours for odd $i$, and $n$th order right neighbours for even $i$ (see Methods). The weak and strong interaction limits of this Hamiltonian provide a simple picture that allows us to rationalise the results presented in Figs.~\,\ref{fig:fig2} and \ref{fig:fig3}.

In the first limit we consider an interaction smaller than the bandwidth $U_i \ll t_1, t_2$, obtaining the standard SSH model\,\cite{ssh1,ssh2}. For $t_2<t_1$, we obtain vanishing Berry phases for electrons, holes, and excitons, with all of them localised at the origin of the unit cell [Fig.\,\ref{fig:fig3}(a)], and corresponding to region I of Fig.\,\ref{fig:fig2}. For $t_2 > t_1$, and in this dispersive limit, the electrons, holes, and as a consequence, the excitons, all acquire $\pi$ Berry phases. This is also reflected by the inversion-symmetry protected shift of all with respect to the unit cell origin [Fig.\,\ref{fig:fig3}(b)], corresponding to region II of Fig.\,\ref{fig:fig2}.

The other limit we consider corresponds to dominant interactions. This limit is characterised by flat bands ($t_2 \gg t_1$ or $t_2 \ll t_1$) and the interactions take the dominant role $U_i \gg t_1, t_2$. We focus on semilocal short-range interactions with $i=1,2$. The electronic hopping terms become negligible and the nearest-neighbour interactions, modelled with the Hubbard terms, take over the role of the quasiparticle hoppings, with density operators taking the role of creation and annihilation operators after a particle-hole transformation $h^{\dagger}_{A/B}= c_{A/B} $, and a bosonisation to the localised exciton basis ${b^{\dagger}_{A/B}= c^{\dagger}_{A/B} h^{\dagger}_{A/B}}$, see Supplementary Information (SI). On relabelling the dominant nearest-neighbour interaction energies $U_1$ and $U_2$ as $2 t^{\text{exc}}_{1}$ and $2 t^{\text{exc}}_{2}$, the dual Hamiltonian takes an SSH form:
\beq{}
\begin{split}
    H = -t^{\text{exc}}_{1} \sum_{j} b^{\dagger}_{B,j} b_{A,j} -t^{\text{exc}}_{2} \sum_{j} b^{\dagger}_{{B+1},j} b_{A,j} + \text{h.c.}.
    \label{eq:dual}
\end{split}
\eeq

Interestingly, this regime with dominant interactions described by the dual Hamiltonian in Eq.\,(\ref{eq:dual}) hosts two new phases which are not present in the phase diagram in Fig.\,\ref{fig:fig2}. When $t^{\text{exc}}_1 > t^{\text{exc}}_2$ and $t_2 \gg t_1$, we obtain a regime in which we have topological electrons and holes with $\phi_{\text{e}} = \phi_{\text{h}} = \pi$  but trivial excitons with $ \phi_{\text{exc}} = 0$. This regime is associated with electron Wannier centres localised at the edges of the unit cells but exciton Wannier centres localised at the centres of the unit cells, see Fig.\,\ref{fig:fig3}(c). When $t^{\text{exc}}_2 > t^{\text{exc}}_1$ and $t_1 \gg t_2$, we obtain a regime in which we have trivial electrons and holes with $\phi_{\text{e}} = \phi_{\text{h}} = 0$  but topological excitons with $ \phi_{\text{exc}} = \pi$. This regime, which was recently described in Ref.\,\cite{davenport2024interactioninduced}, is associated with electron Wannier centres localised at the centres of the unit cells but exciton Wannier centres localised at the edges of the unit cells, see Fig.\,\ref{fig:fig3}(d). Both of these regimes exhibit exciton phases in which the topology is either driven or suppressed by the electron-hole interactions rather than by the underlying single-particle states.

A natural question to ask at this point is why the two interaction-dominated regimes realisable in the SSH-Hubbard model~\cite{davenport2024interactioninduced} are missing from Fig.\,\ref{fig:fig2}. To understand this, we note that in the SSH-Hubbard model~\cite{davenport2024interactioninduced} the single-particle parameters $t_1$ and $t_2$ and the interaction parameters $U_1$ and $U_2$ are taken as independent parameters. However, in a more realistic setting, such as that provided by the Wannier equation used to build the phase diagram in Fig.\,\ref{fig:fig2}, they are not independent. As such, our numerical results obtained from the Wannier equation show that, for the one-dimensional systems studied, excitons only exhibit topology inherited from the underlying single-particle electron and hole topologies. An interesting question to explore in future work would be the possibility of interaction-driven topology in excitons, perhaps in higher-dimensional settings. 

\subsection*{Topological excitons in polyacenes}
Most importantly, we identify a family of organic acene one-dimensional polymers\,\cite{cirera2020tailoring,PhysRevB.106.155122} as candidate materials to realise the predicted exciton topological phase described above. These materials are formed of $N$-ring acene molecules linked by a carbon-carbon bond on the central carbon atoms (see Fig.\,\ref{fig:fig3} for a schematic), and we consider the cases $N=3,5,7$.

We use density functional theory\,\cite{PhysRev.136.B864, RevModPhys.87.897, Giannozzi_2009, Giannozzi_2017} to evaluate the single-particle states of these polyacenes, and then fit the results to the SSH model to identify effective $t_1$ and $t_2$ hopping parameters. We indicate the location of polyanthracene ($N=3$),  polypentacene ($N=5$) and polyheptacene ($N=7$) in the phase diagram of Fig.\,\ref{fig:fig2}(a), providing material realisations of both trivial and topological excitonic phases. 

The excitonic topology of polyacenes can be controlled through the application of strain along the polymer chains, as illustrated in Fig.\,\ref{fig:fig2}(b). Unstrained polyanthracene sits in region I, and by applying a tensile ($+5$\%) or compressive ($-5$\%) strain the bandstructure can be made less or more dispersive, respectively. In the case of polypentacene, which sits in region II for the unstrained case, strain has the opposite effect on the evolution of the band dispersion due to the topological electronic band inversion with respect to the polyanthracene case. 
Interestingly, unstrained polypentacene sits sufficiently close to the trivial-topological phase boundary that tensile strain could potentially drive a topological phase transition in this compound. From our density functional theory (DFT) calculations (see Methods), we find that this transition occurs between $+5$\% and $+10$\% strain, demonstrating that the topology of excitons can be controlled via strain, see also Supplementary Fig.~3.
More generally, it would be interesting to explore how topology can be manipulated by other external parameters besides strain, for example through temperature or external electromagnetic fields~\cite{thompson2024topologicallyenhancedexcitontransport}.

\begin{figure} [!t]
    \centering
    \includegraphics[width=0.8\linewidth]{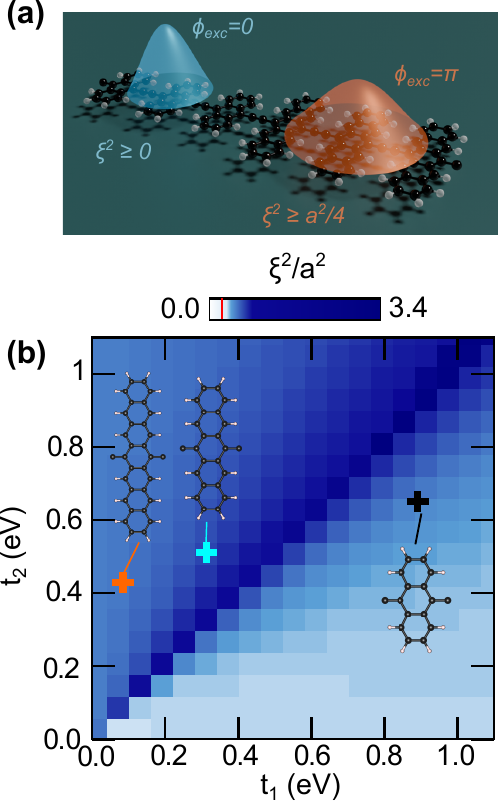}
    \caption{{\bf Quantum geometry of topological excitons.} \textbf{(a)} Illustration of topological (orange) and trivial (blue) excitons in polyacenes with the corresponding excitonic Berry phases and excitonic Wannier state spreads bounded below by the excitonic quantum metric. \textbf{(b)} Spread of the maximally-localised excitonic Wannier states $\xi^2$ as a function of the single-particle hopping parameters $t_1$ and $t_2$, with the excitonic phase bounded by the excitonic quantum geometry (red line in gradient bar). The corresponding parameters and $\xi^2$ values for polyheptacene (orange cross), polypentacene (cyan cross), and polyanthracene (black cross) are indicated.}
    \label{fig:fig4}
\end{figure}

\subsection*{Excitonic quantum geometry} 
The excitonic topology described above has a direct implication on a geometric property, the spread (variance) of the excitonic states in the centre-of-mass coordinate $\vec{R}$. To understand this relation, we consider the quantum metric $g^{\text{exc}}_{ij}$\,\cite{provost1980riemannian,Resta_2011}, which is a tensor made of symmetrised derivatives of momentum-space Bloch states, and defined as:
\beq{}
    g^{\text{exc}}_{ij} = \frac{1}{2} \Big[ \bra{\partial_{Q_i} u^{\text{exc}}_\textbf{Q}} 1 - \hat{P} \ket{\partial_{Q_j} u^{\text{exc}}_\textbf{Q}} + \text{c.c.} \Big],
\eeq
where $\hat{P} = \ket{u^{\text{exc}}_\textbf{Q}}\bra{u^{\text{exc}}_\textbf{Q}}$ is a projector onto the excitonic band of interest. More generally, it can be considered to be the real part of a Hermitian quantum geometric tensor (QGT)\,\cite{provost1980riemannian,tormaessay}, which can be written as\,\cite{provost1980riemannian,bouhon2023quantum}:
\beq{}
    Q^{\text{exc}}_{ij} = \bra{\partial_{Q_i} u^{\text{exc}}_\textbf{Q}} 1 - \hat{P} \ket{\partial_{Q_j} u^{\text{exc}}_\textbf{Q}}.
\eeq
The imaginary part of the QGT corresponds to the Berry curvature, encoding the topology that underlies quantum Hall phenomena, while the positive-semidefiniteness of the QGT equips the metric and the Berry curvature with a set of physical bounds related to optics, superconductivity, and quantum transport\,\cite{tormaessay}. {As we show in the following, the QGT directly determines the spread of the excitonic states. The geometric meaning of its real part, the metric, can be further identified} by defining a Fubini-Study metric\,\cite{provost1980riemannian}, ${\dd s^2 = 1 - \Big| \bra{u^{\text{exc}}_\vec{Q}} \ket{u^{\text{exc}}_{\vec{Q}+\dd \vec{Q}} }\Big|^2}$, and recognizing that:
\beq{}
    \dd s^2 = g^{\text{exc}}_{ij} \dd Q^i \dd Q^j,
\eeq
which is reminiscent of the well-known relation between the metric and the spacetime intervals in general relativity, and where the Einstein summation convention is implicitly assumed.

Focusing on one spatial dimension, which we refer to as the $x$ direction, the metric has a single component $g^{\text{exc}}_{xx}$. The metric is related to the exciton variance $\xi^2\equiv \text{Var}\,R = \langle R^2 \rangle - \langle R\rangle^2$ of the centre of the maximally-localised exciton Wannier functions (MLXWF)\,\cite{PhysRevB.108.125118} in real space. Explicitly, for a unit cell of length $a$, we have $\xi^2=\frac{a}{2\pi}\int_{\mathrm{BZ}} \dd Q\,g^{\text{exc}}_{xx}$. Using a Cauchy-Schwarz type inequality, we find (see Methods): 
\beq{}
 \xi^2 \geq \frac{a^2 P_{\text{exc}}^2}{4}.
\eeq
This relation explicitly shows that the size of topological excitons, as captured by the spread $\xi^2$ of the excitonic Wannier functions, should be comparable to, or exceed, the size of the unit cell. That is, topological excitons with $P_{\text{exc}}=1$ are shifted to the boundary of the unit cell $\langle R \rangle = a/2$ and their spread $\xi^2$ is bounded from below, resulting in the centre-of-mass $R$ of an exciton being localised within a characteristic length $\xi \sim a/2$. In the presence of a localised electron or hole (e.g. through defect pinning), the bound on the delocalisation of the centre-of-mass $R$ position also determines the effective size of the exciton characterised by the electron-hole distance $r$. This is consistent with the observed transition from localised ``Frenkel'' to spread out ``Mott-Wannier'' excitons in polyacenes\,\cite{PhysRevB.106.155122}.

To illustrate the theoretical results on the lower bound on the exciton centre-of-mass spread, we numerically demonstrate the spread of the excitons in Fig.\,\ref{fig:fig4}. We confirm that whenever the topological excitonic invariant $P_{\text{exc}}$ is non-trivial, the diameter associated with the variance of the real space exciton wavefunctions exceeds the size of the unit cell. By contrast, Fig.\,\ref{fig:fig4} also shows that for an excitonic inversion invariant $P_{\text{exc}}=0$, the excitons have no lower bound on their spread and they are more localised compared to their topological counterparts.

Finally, we find that increasing the dielectric screening can be used to fine tune the exciton spread, while satisfying the topological bound. In an experimental setting, the screening strength can be controlled through substrate choice, but also through chemical functionalisation. More details can be found in the SI.

\begin{figure*} [!t]
    \centering
    \includegraphics[width=0.9\linewidth]{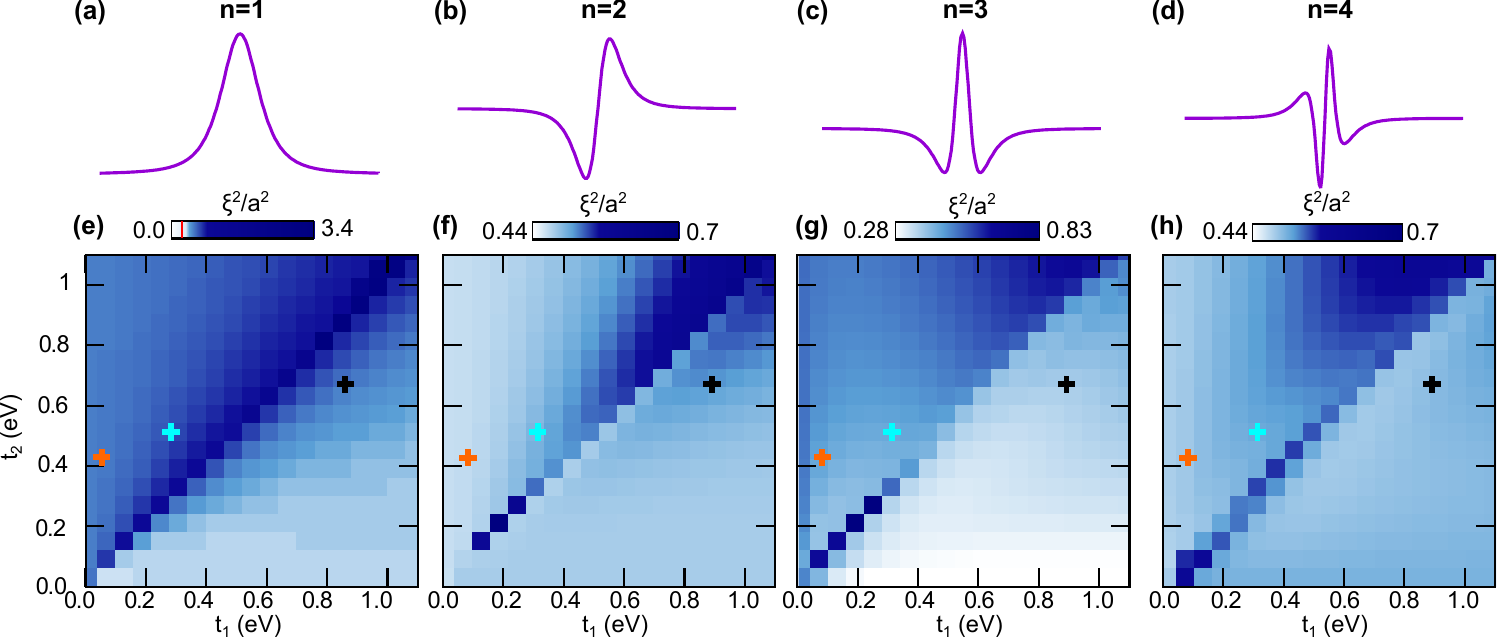}
    \caption{{\bf Topological excitons in higher excitonic bands.} \textbf{(a-d)} Envelope functions $\psi_{\textbf{Q}=0}(\kv)$ for different excitonic bands $n = 1,2,3,4$, respectively. \textbf{(e-h)} Spread of the excitonic Wannier states $\xi^2$, with the excitonic topological phases bounded by the excitonic quantum geometry (red line in gradient bar). Although the numerical results match the general theory and topological phase diagram intimately, we note that in higher excitonic bands $n=3,4$ \textbf{(g-h)}, there is a larger spread $\xi^2$ than in the topologically trivial excitonic wavefunctions of lower bands \textbf{(e-f)}, due to the contributions of more delocalised envelope function (see also SI). 
    The excitons in polypentacene (cyan cross), polyheptacene (black cross) and polyheptacene (orange cross) realise different excitonic topologies.}
    \label{fig:fig5}
\end{figure*}

\subsection*{Higher-energy excitonic bands} 
The discussion up to this point has focused on the lowest-energy excitonic band. This is typically the most interesting excitonic band from an optoelectronics point of view, but the discussion above applies equally to higher energy excitonic bands. To illustrate this, Fig.\,\ref{fig:fig5}(a)-(d) illustrates the exciton envelope functions for excitonic bands $n=1,2,3,4$ while the associated  spread of the exciton Wannier states are shown in  Fig.\,\ref{fig:fig5}(e)-(h). The topological phase diagram for each of these states is the same owing to the vanishing envelope contribution to the excitonic Berry connection~\cite{Kwan2021}.

The quantum geometric bound driven by the excitonic band topology persists in the higher excitonic bands, see Fig.\,\ref{fig:fig5}(e)-(h). In the higher bands, quantum geometry takes a more dominant role, even without the presence of non-trivial topology, due to the broader envelope functions of the weaker-bound higher energy excitonic states (see further discussion of this point in SI).

\section*{Discussion}
Our findings open up multiple research avenues. From a theoretical point of view, the SSH-like topology discussed above, also referred to as obstructed insulators, is the simplest kind of band topology. Extending the study of topological excitons to higher dimensional systems and to materials with different types of (crystalline) symmetries is expected to greatly expand excitonic topology and quantum geometry\,\cite{Kwan2021,Brey2021,MaiselLiceran2023,bouhon2023quantum,davenport2024interactioninduced}. Another interesting research direction is to extend the discussion above to spinful models, which would enable a distinction between topological properties of singlet and triplet excitons, which in this work we find to be qualitatively the same, in a broader range of materials.

These potential theoretical extensions would naturally fit with active areas of experimental exciton research. Excitons are experimentally studied through their optical, dynamics, and transport properties, using techniques such as pump-probe experiments. The findings reported above already anticipate some potential experimental manifestations of excitonic topology. First, the strain-controllable exciton topologies and geometries exhibited by different exciton states (see {Figs.\,\ref{fig:fig2}, \ref{fig:fig5}}) suggest that optical probes could selectively target topologically-distinct excitons under different external conditions applied to the same material, as well as unusual dynamics as these excitons relax towards the lowest energy states following photoexcitation. Second, the lower bound set on the exciton size by quantum geometry (see Fig.\,\ref{fig:fig4}) suggests that transport will be enhanced for topological excitons compared to their trivial counterparts, an intriguing possibility we have very recently explored in Ref.\,\cite{thompson2024topologicallyenhancedexcitontransport}. Third, a hallmark of topology is the so-called bulk-boundary correspondence, in which the bulk topology is associated with unusual boundary states that are often protected against scattering and support dissipationless transport in higher dimensions. In the case of topological excitons, such topological excitonic boundary states should be experimentally observable in local optical conductivity measurements, as suggested by Ref.\,\cite{davenport2024interactioninduced}.

From a materials point of view, the interplay between topology, excitons, and organic polyacenes reported above is only the starting point. Organic semiconductors provide a more general platform to explore excitonic topology, with organic semiconductors in one, two, and three dimensions often exhibiting strongly bound excitons, due to their weak dielectric screening. Additionally, it should be possible to harness the decades of experience in organic chemistry to manipulate organic semiconductors to explore many potential different exciton topological regimes, such as those described in Fig.\,\ref{fig:fig3}, but notably also richer topologies in higher dimensions. Beyond organic semiconductors, inorganic two-dimensional materials \cite{perea2022exciton, MaiselLiceran2023} and van der Waals bonded layered materials \cite{alexeev2019resonantly, jin2019observation, dyksik2021brightening} also exhibit strongly bound excitons, providing additional material platforms to explore excitonic (crystalline) topology.

Overall, we establish a connection between topological physics, common in the study of electronic phases, and the field of excitons in organic semiconductors. We describe a simple proof-of-principle for excitonic topology in the form of one dimensional crystalline topological invariants, constructing a full excitonic topological phase diagram with different regimes reflecting the interplay between the underlying (so-called obstructed) electron and hole topologies with the new excitonic topology. We also present a family of organic polymers that host the predicted exciton topologies, and demonstrate the manipulation of this topology by applying experimentally-realisable strains. Finally, we discover that excitonic topology is related to the spatial localisation of excitons, as determined by Riemannian metric identities. In particular, we find a lower bound on the exciton size for topologically non-trivial excitons. We show that the associated, topologically-bounded, excitonic quantum geometry can be further controlled with dielectric screening through a substrate choice, see Supplementary Fig.~2. Our results set a benchmark for a potentially rich exploration of topological excitons in organic semiconductors and beyond, which has the potential to impact properties ranging from optical to extraordinary transport signatures.

\section*{Methods}
\subsection*{First principles calculations}
We perform density functional theory calculations using the {\sc Quantum Espresso} package\,\cite{Giannozzi_2009,Giannozzi_2017} to study a family of polyacene chains. We use kinetic energy cutoffs of $80$\,Ry and $500$\,Ry for the wavefunction and charge density, respectively, and $12$ $k$-points to sample the Brillouin zone along the periodic direction. We use GGA (PBE) norm-conserving pseudopotentials which were generated using the code {\sc oncvpsp} (Optimized Norm-Conserving Vanderbilt PSeudoPotential)\,\cite{hamann2013optimized} and can be found online via the Schlipf-Gygi norm-conserving pseudopotential library\,\cite{schlipf2015optimization}. We impose a vacuum spacing of $34.3$\,\AA\ in the planar direction perpendicular to the polymer chain, and a vacuum spacing of $27.52$\,\AA\ in the out-of-plane direction, to minimise interactions between periodic images of the polyacene chains. We perform a structural optimisation of the internal atomic coordinates to reduce the forces below $0.0015$\,Ry/\AA.  To model the impact of strain we expand/compress the unit cell in the polymer chain direction before relaxing the atomic positions. For further details concerning atomic positions, see Supplementary Data 1.

\begin{figure*} [!t]
    \centering
    \includegraphics[width=\textwidth]{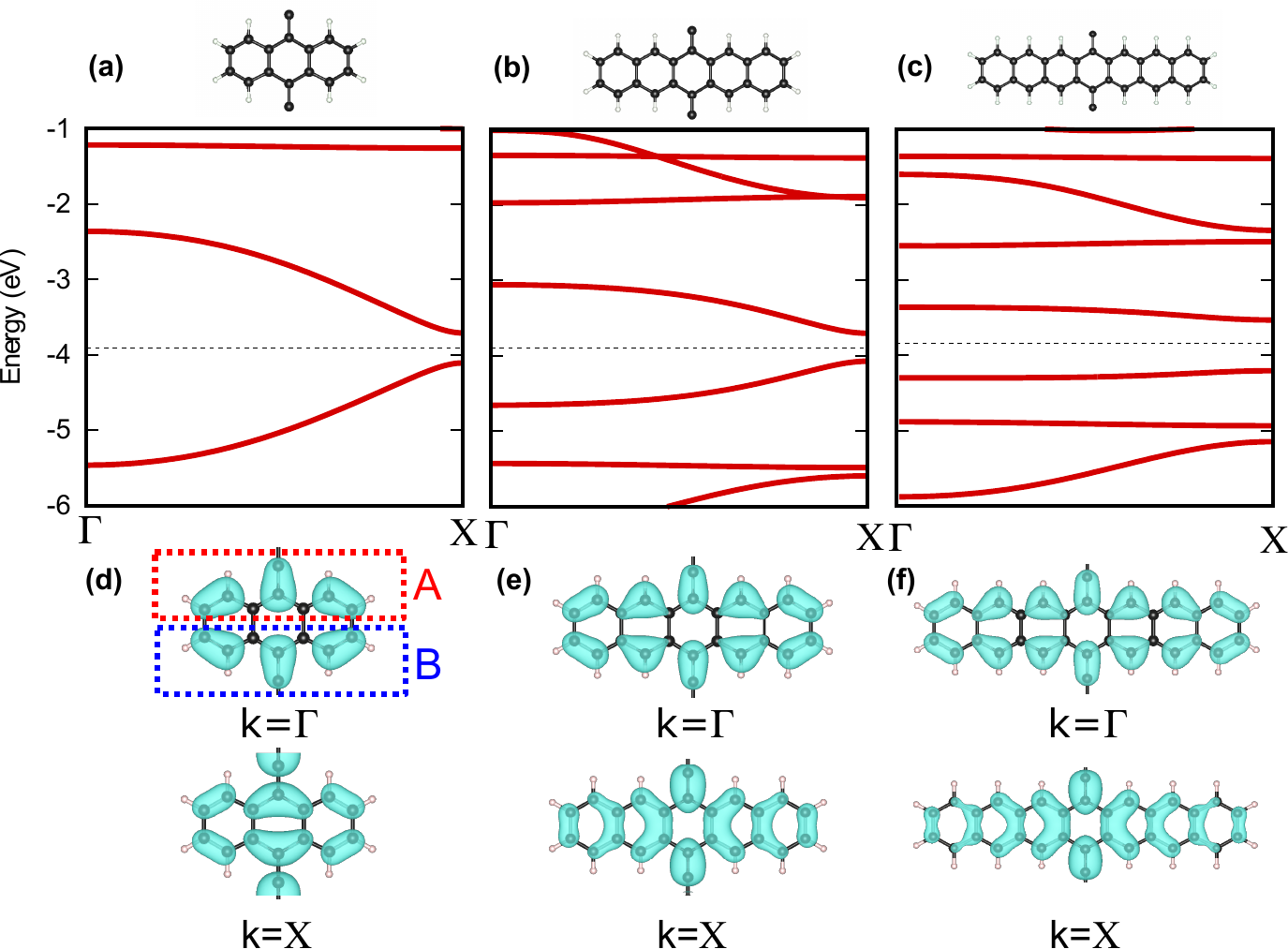}
    \caption{{\bf Electronic band topology in organic semiconductors.} Electronic bands in \textbf{(a)} polyanthracene, \textbf{(b)} polypentacene, and \textbf{(c)} polyheptacene, calculated using DFT. The corresponding Fermi levels are indicated with black dashed lines. \textbf{(d-f)} Plots of the electronic wavefunctions for the valence band in the polyacenes, which is contributed by the $p_z$-orbitals. The shift of the charge centre, as well as the change of the parity of the wavefunction, as indicated by the inversion symmetry eigenvalues at the high symmetry points of a one-dimensional momentum space, $k = \Gamma$ (BZ centre), and $k = \text{X}$ (BZ edge), can be directly observed (on comparing $P_{\text{exc}} = 0$ to $P_{\text{exc}} = 1$) in the DFT results. The approximate $A$ and $B$ sublattice sites in the molecular unit cells are marked with the red and blue boxes, respectively.}
    \label{fig:figS1}
\end{figure*}

We calculate the band structures of several polyacenes, as shown in Fig.\,\ref{fig:figS1}. The hopping parameters of the SSH tight-binding model were deduced from these calculations. The SSH sites, $A$ and $B$, on the molecular unit cell are illustrated with the red and blue boxes in Fig.\,\ref{fig:figS1}(d), with the $t_1$ hopping across an acene molecule and the $t_2$ hopping across the carbon link between acenes\,\cite{cirera2020tailoring}. 

Using these calculations, we identify polyanthracene ($N=3$) as exhibiting topologically trivial electronic states, while polypentancene ($N=5$) and polyheptacene ($N=7$) exhibit topological electronic states. Repeating the same analysis using the many-body $GW$ approximation to calculate the band structure of the polyacenes leads to the same topological transition as the polyacene chain length increases as that calculated at the DFT level, but with the transition occurring for longer chain lengths\,\cite{PhysRevB.106.155122}. Specifically, the transition from trivial to topological electrons and holes occurs between the polypentacene ($N=5$) and polyheptacene ($N=7$) polymers, in contrast with our and previous\,DFT calculations and experiments \cite{cirera2020tailoring}  where the transition occurs between $N=3$ and $N=5$.

\begin{figure*} [!t]
    \centering
    \includegraphics[width=\textwidth]{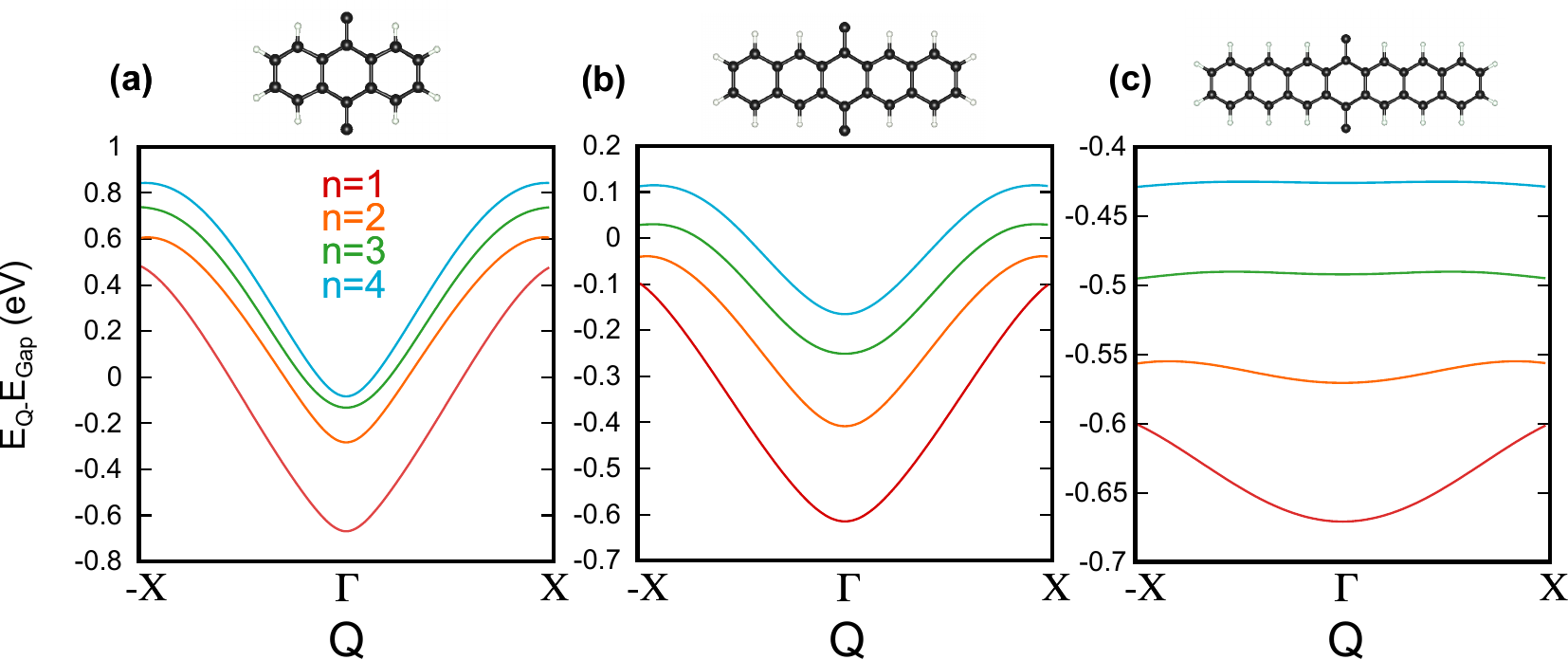}
    \caption{{\bf Excitonic bands in organic polymers.} Excitonic bands obtained from DFT single-particle states and the solution of the Wannier equation for the envelope function. We show the dispersion of the excitonic bands $n=1$ (red), $n=2$ (orange), $n=3$ (green), and $n=4$ (blue) of \textbf{(a)}~polyanthracene ($N=3$), \textbf{(b)} polypentacene ($N=5$), and \textbf{(c)} polyheptacene ($N=7$). The excitonic bands are manifestly inversion-symmetric, while the excitonic bandwidths and dispersion decrease with increasing number of rings $N$ in the monomers. We also note that as the excitonic bands approach the zero energy level, they merge into the continuum of electronic energies. 
    }
    
    \label{fig:figS2}
\end{figure*}

\subsection*{Wannier equation}

We obtain the excitonic wavefunctions and associated bands by combining the electronic and hole states from the SSH model or from the first principles DFT calculations, with the envelope part of the exciton wavefunction obtained by solving the Wannier equation. For the illustrations of the envelope part solutions, see also Supplementary Fig.~1.

For each pair ($t_1, t_2$), we solve the Wannier equation:
\beq{}
    \sum_{k'} h_{k,k'}(Q) \psi^n_{Q}(k) = E_n(Q) \psi^n_{Q}(k),
\eeq
where $\psi^n_{Q}(k)$ is the envelope function of the $n$th excitonic band. In this expression, we have:
\begin{align}
  h_{k,k'}(Q) = (E^e_{k+Q/2} -  E^h_{k-Q/2} ) \delta_{k, k'} - W_{k,-k', Q}, 
\end{align}
with electron/hole band energies $E^{e(h)} = E_{+(-)}$, and the interaction matrix defined as~\cite{MaiselLiceran2023}:

\begin{align*}
   W_{k,-k', Q} &= V_\text{NR}(k-k')\\& \times 
   \sum_{i,j \in\{A,B\}} \varphi^*_{i,k+Q/2} \varphi^*_{j,k'-Q/2}
   \varphi_{j,k-Q/2}\varphi_{i, k'+Q/2},
\end{align*}

where $\varphi_{i,k}$ is the wavefunction amplitude from the SSH model on site $i$ with momentum $k$, and $V_{\text{NR}}$ is the Coulomb potential for a one-dimensional system that describes the dielectric screening.
We show the excitonic band dispersions for the lowest four excitonic bands of different polyacenes in Fig.\,\ref{fig:figS2}.

The Wannier equation is highly dependent on the dielectric screening, where an accurate screening model is paramount to adequately capture the excitonic physics. In the polyacene chains, the Coulomb potential corresponds to that of one-dimensional nanoribbons. Hence, consistently with Ref.\,\cite{villegas2024screened}, on solving Poisson's equation, we arrive at the following form of the Coulomb potential: 

\begin{align}
    V_\text{NR}(Q) = \dfrac{e_0^2}{4\pi\epsilon_0} \dfrac{K_0(Q L/2)}{\varepsilon_s+8Q^2\alpha_{\text{1D}}K_0(Q L/2)},
\end{align}
where $L$ is the width of the ribbon, $\alpha_{\text{1D}}$ is the screening parameter (polarisability per unit length), $\varepsilon_s$ is the background screening, and $K_0$ is a modified Bessel function of the second kind. We take $L$ to be the lateral size of the polyacene ribbon and fix $\alpha_{\text{1D}}=0.05$\,nm$^{-2}$ following previous work\,\cite{villegas2024screened}. For small $Q$, corresponding to long range interactions, the screening is dominated by the environment as determined by $V_\text{NR}(Q\rightarrow 0) \sim -\log(QL/2)/\varepsilon_s $. At large $Q$, corresponding to short distances, the screening is approximately $V_\text{NR}(Q\rightarrow 0) \sim 1/(\alpha_{\text{1D}}Q^2)
$, resembling the  Coulomb interaction in bulk systems. The background screening $\varepsilon_s$,  defined as the average of the dielectric above and below the organic layer, further modulates the Coulomb interaction. We consider common background dielectrics in this work: vacuum (${\varepsilon_s = 1}$), SiO$_2$ substrate (${\varepsilon_s = 2.45}$), and hBN encapsulation (${\varepsilon_s = 4.5}$)\,\cite{thompson2022anisotropic}. 

\subsection*{Numerical excitonic Berry phases}
To evaluate the excitonic Berry phases, we use a discretisation of the Brillouin zone $Q = \left[ Q(1), Q(2), \ldots, Q(N_Q) \right]$, where we choose $Q(1) = 0$ and $Q(N_Q) = 2\pi/a-\Delta Q$, with lattice parameter $a$ and grid spacing $\Delta Q = 2\pi/(N_Q a)$. We use the parallel-transport gauge to numerically evaluate the Berry phase given by Eq.\,\eqref{eq::excBerry}, following Ref.\,\cite{vanderbilt2018berry}:
\beq{}
    \phi_\text{exc} = -\mathfrak{Im}~\text{log} \Big[ \bra{u^{\text{exc}}_{Q(1)}}
    \ket{u^{\text{exc}}_{Q(2)}} \ldots \bra{u^{\text{exc}}_{Q(N_Q-1)}}\ket{u^{\text{exc}}_{Q(N_Q)}} \Big]. 
\eeq
We use a $Q$-grid of size $N_Q=400$, with convergence within a numerical error of $\Delta \phi_{\text{exc}} \sim 10^{-3}$ for the excitonic Berry phase obtained above $N_Q=200$.

\begin{figure} [!h]
    \centering
    \includegraphics[width=\linewidth]{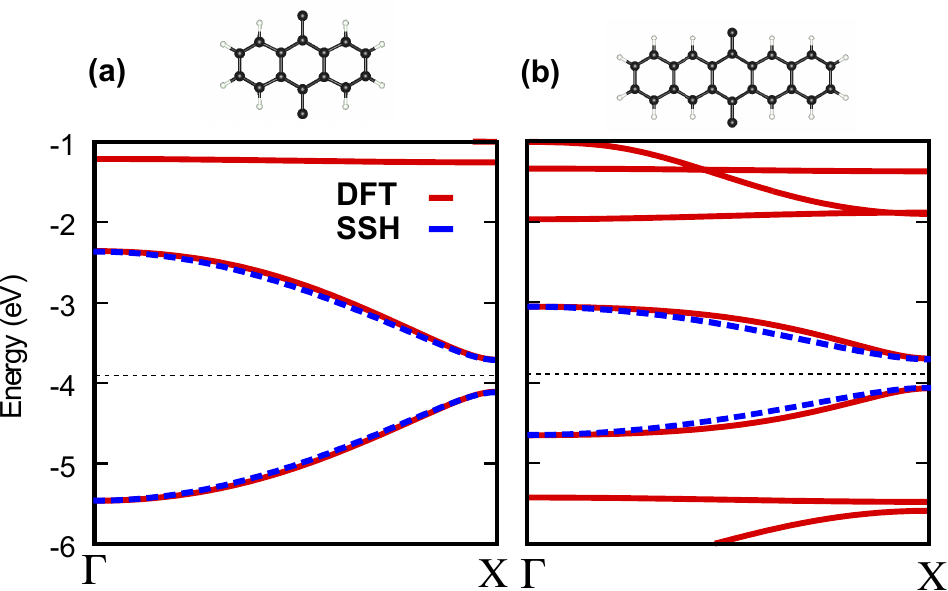}
    \caption{{\bf DFT bands and the electronic SSH model.} Low energy band fitting of the effective SSH model (blue) to the band structure obtained with DFT calculations (red) for {\bf (a)} polyanthracene and {\bf (b)} polypentacene. The black dashed lines indicate the Fermi levels. We note that apart from the energy fitting, the form of the electronic wavefunction and its parity, as provided in Fig.~\ref{fig:figS1} and consistent with Ref.~\cite{cirera2020tailoring}, are necessary to show the topological nature of the electronic states that further contribute to the formation of topological excitons.}
    \label{fig:dft-ssh-2}
\end{figure}

\subsection*{SSH-Hubbard model}

In this Section we expand on the effective model discussed in the main text. We begin by building an electronic effective theoretical model based on the one-dimensional SSH Hamiltonian\,\cite{ssh1,ssh2}. We write Eq.~\eqref{eq::elecSSH} of the main text in the momentum-space Bloch orbital basis by introducing $c^\dagger_{Ak}/c_{Bk}$, on Fourier transforming the electronic creation and annihilation operators in the position basis $c^\dagger_{Ak} = \frac{1}{\sqrt{L}} \sum_{j=1} e^{\ii k (r_{\mathrm{e}})_j} c^\dagger_{A,j}$,
\beq{}
H = \sum_{k; \alpha,\beta = A,B} \left[H^{\text{SSH}}(k)\right]_{\alpha \beta} c^\dagger_{\alpha k} c_{\beta k},
\eeq
with the electron position operator $(r_{\mathrm{e}})_j = ja$. In this basis, we have an effective Bloch Hamiltonian matrix for low energy, i.e. top valence and bottom conduction, bands, which reads:
\begin{equation}
H^{\text{SSH}}(k) = \vec{d}(k) \cdot \boldsymbol{\sigma},    
\end{equation}
where $\vec{d}(k) = (t_1 + t_2 \cos k, t_2 \sin k, 0)^T$, with $\boldsymbol{\sigma} = (\sigma_x, \sigma_y, \sigma_z)$, the vector of Pauli matrices. The hopping parameters $t_1$ and $t_2$ can be retrieved to describe specific polyacenes by fitting them to the associated first principles calculations after Wannierisation. In this electronic model, the wavefunctions of electrons and holes take the compact form ${\ket{u^{\text{e}/\text{h}}_k} = \frac{1}{\sqrt{2|\vec{d}|^2}} ({d_x} \pm \ii {d_y}, |d|)^\text{T}}$. Correspondingly, we further demonstrate the quantitative fit of the discussed electronic SSH model to the first principles band structures; see Fig.~\ref{fig:dft-ssh-2} below.

Having set up the electronic problem, we now elaborate on the theoretical model for the phenomenology of the interaction-dependent excitonic topology. We start with a SSH-Hubbard-type model with up to the $n$th neighbor interactions $U_i$, with $i = 1, 2, 3, \ldots, n$ in real space:
\\
\beq{}
\begin{aligned}
    H &= \sum_{j} \epsilon_A n_{A,j} + \sum_{j} \epsilon_B n_{B,j}  \\&-t_1 \sum_{j} c^{\dagger}_{A,j} c_{B,j} + \text{h.c.} -t_2 \sum_{j} c^{\dagger}_{A,j+1} c_{B,j} + \text{h.c.} \notag\
    \\&+ \sum_{j} U_A n_{A,j} (n_{A,j} - 1)
    + \sum_{j} U_B n_{B,j} (n_{B,j} - 1) \\&+ \frac{1}{2} \sum_{j} \sum_{\alpha,\beta = A,B} \sum^{n}_{i=1} \Big[ U^{\alpha \beta}_{2i-1}  n_{\alpha,j-i} n_{\beta,j} + U^{\alpha \beta}_{2i} n_{\alpha,j+i} n_{\beta,j} \Big],
\end{aligned}
\eeq

where the density operator is ${n_{A,j} = c^{\dagger}_{A,j} c_{A,j}}$. We set the onsite energies to zero, ${\epsilon_A = \epsilon_B = 0}$, and the onsite interaction terms of strength $U_A$, $U_B$ are further set to vanish within a spinless model. Consistently, the model admits no self-interactions. We additionally note that if the screened electron-electron interaction potential takes a Coulomb-like decaying form, ${V(r - r') \sim \frac{1}{\varepsilon_s |r-r'|}}$, the interaction strengths can be completely suppressed as $U_1, U_2 \sim 1/\varepsilon_s \rightarrow 0$, when formally, $\varepsilon_s \rightarrow \infty$. Equivalently, this limit yields a hard-core repulsive potential $V(r - r') \sim \delta(r-r')$, where $\delta(r-r')$ is the Dirac delta function.

In the real material context, and consistently with our fitting, we stress that all the phenomenological model parameters $t_1, t_2, U_i, \text{etc.}$, are self-contained in the self-consistent solution of the Wannier equation, but otherwise these can be directly estimated to an arbitrary order from the electronic Wannier functions. For the standard computation of the Wannier functions using {\sc Wannier90}, see Ref.~\cite{MOSTOFI20142309}.

\section*{Data availability}

All datasets for the plots of this study are available upon request to the authors.

\section*{Code availability}
All codes and associated data are reproducible with information in the manuscript. All first-principles calculation input files are available upon request to the authors.

\section*{Acknowledgements}
The authors thank Richard Friend, Akshay Rao, Sun-Woo Kim, Gaurav Chaudhary, Arjun Ashoka, Henry Davenport, and Frank Schindler for helpful discussions. This project was supported by funding from the Rod Smallwood Studentship at Trinity College, Cambridge (W.J.J.). We acknowledge support from a EPSRC Programme grant EP/W017091/1 (J.J.P.T.,~and~B.M.), as well as from UKRI Future Leaders Fellowship MR/V023926/1, from the Gianna Angelopoulos Programme for Science, Technology, and Innovation, and from the Winton Programme for the Physics of Sustainability (B.M.). We acknowledge funding from a New Investigator Award, EPSRC grant EP/W00187X/1, a EPSRC ERC underwrite grant EP/X025829/1, and a Royal Society exchange grant IES/R1/221060, as well as from Trinity College, Cambridge (R.-J.S.).

\bibliographystyle{naturemag}
\bibliography{references}

\section*{Author Contributions Statement} B.M. and R.-J.S. initiated the project. W.J.J. performed initial theoretical analysis of excitonic topology and geometry with inputs from R.-J.S. J.J.P.T. performed all numerical first-principles calculations with inputs from B.M. W.J.J. and R.-J.S. constructed theoretical phenomenological model with inputs from J.J.P.T. and B.M. All authors discussed the results and substantially contributed to the writing of the manuscript. The final form of the manuscript, including Methods and Supplementary Information, benefitted from input from all authors. 

\section*{Competing Interests Statement} The authors declare no competing interests.



\newpage

\newpage
\clearpage

\end{document}


\preprint{APS/123-QED}

\title{Supplementary Information for ``Excitonic topology and quantum geometry in organic semiconductors"}

\author{Wojciech J. Jankowski}
\email{wjj25@cam.ac.uk}
\thanks{}
\affiliation{TCM Group, Cavendish Laboratory, Department of Physics, J J Thomson Avenue, Cambridge CB3 0HE, United Kingdom}

\author{Joshua J.P. Thompson}
\thanks{}
\affiliation{Department of Materials Science and Metallurgy, University of Cambridge,
27 Charles Babbage Road, Cambridge CB3 0FS, United Kingdom}

\author{Bartomeu Monserrat}
\thanks{}
\affiliation{TCM Group, Cavendish Laboratory, Department of Physics, J J Thomson Avenue, Cambridge CB3 0HE, United Kingdom}
\affiliation{Department of Materials Science and Metallurgy, University of Cambridge,
27 Charles Babbage Road, Cambridge CB3 0FS, United Kingdom}

\author{Robert-Jan Slager}
\email{rjs269@cam.ac.uk}
\affiliation{TCM Group, Cavendish Laboratory, Department of Physics, J J Thomson Avenue, Cambridge CB3 0HE, United Kingdom}

\date{\today}

\maketitle 

In this Supplementary Information, we provide further details on:\\

\begin{enumerate}[leftmargin=47.5mm, label=
 \textbf{Supplementary Note \arabic*:}]
    \item Wannier centres and Berry phases
    \item Decomposition of the excitonic Berry connection
    \item Dualisation of the interacting Hamiltonian
    \item Controlling excitonic topology with strain
    \item Riemannian geometry of excitons
    \item Derivation of the bound on excitonic spread
    \item Numerical excitonic localisation
\end{enumerate}

\section*{Supplementary Note 1: Wannier centres and Berry phases}

We here briefly comment on the interplay of Berry phases and Wannier centres of electrons, holes, and excitons. In the context of electron and hole Wannier centres, we can recapture these in terms of corresponding Wannier centres, $\bar{w}_{e}$ and  $\bar{w}_{h}$. Here, the electron/hole Wannier centres are defined as\,\cite{vanderbilt2018berry}:
\beq{}
   \bar{w}_{e/h} = \bra{w^{e/h}_0} r_{e/h} \ket{w^{e/h}_0},
\eeq
with $r_{e/h}$ the electron/hole position operator, and the corresponding Wannier states $\ket{w^{e/h}_0}$ obtained by Fourier-transforming electron and hole bands\,\cite{vanderbilt2018berry}:
\beq{}
   \ket{w^{e/h}_0} = \frac{a}{2\pi} \int_{\text{BZ}} \dd k~ e^{\ii k r_{e/h}} \ket{u^{e/h}_{k}},
\eeq
over the momentum space coordinates $k$ in the first Brilouin zone (BZ). On the other hand, the excitonic Wannier centres $\bar{w}_{\text{exc}}$: 
\beq{}
   \bar{w}_{\text{exc}} = \bra{w^{\text{exc}}_0} R \ket{w^{\text{exc}}_0},
\eeq
are defined with the exciton centre-of-mass position operator $R$ and exciton Wannier states $\ket{w^{\text{exc}}_0}$:
\beq{}
   \ket{w^{\text{exc}}_0} = \frac{a}{2\pi} \int_{\text{BZ}} \dd Q~ e^{\ii Q R} \ket{u^{\text{exc}}_{Q}},
\eeq
which are also pictorially represented in Fig.~3 of the main text. Additionally, the relation between the Berry phases and the Wannier centres gives $\phi_{e/h} = 2\pi \bar{w}_{e/h}/a$~\cite{vanderbilt2018berry}, which obtains the Berry phases for electron and holes. Analogously, the Berry phase of the excitons is equivalent to the shift of an excitonic Wannier centre $\phi_{\text{exc}} = 2\pi \bar{w}_{\text{exc}}/a$. 

\section*{Supplementary Note 2: Decomposition of the excitonic Berry connection}

To compute the excitonic invariant $P_{\text{exc}}$, we utilise Eq.~(3) of the main text, and decompose the excitonic Berry connection of the Eq.~(2) of the main text into an envelope part and a single-particle part as:
\beq{}
    A_{\text{exc}}(Q) = A_{\text{exc}}^{\text{en}}(Q) + A_{\text{exc}}^{\text{sp}}(Q).
\eeq
The envelope part reads:
\beq{}
    A_{\text{exc}}^{\text{en}}(Q) = \ii \sum_{k} \psi^{\ast}_Q(k) \partial_Q \psi_Q(k),
\eeq
and the single-particle part consists of the electron and hole contributions~\cite{Kwan2021}:
\beq{}
    A_{\text{exc}}^{\text{sp}}(Q) = \frac{1}{2} \sum_{k} |\psi_Q(k)|^2 \left[ A^{\text{e}}(k + Q/2) + A^{\text{h}}(k - Q/2) \right].
\eeq
In turn, the single-particle electron and hole Berry connections are given by $A^{\text{e}/\text{h}}(k \pm Q/2) = \ii \bra{u^{\text{e}/\text{h}}_{k \pm Q/2}} \ket{\partial_Q u^{\text{e}/\text{h}}_{k \pm Q/2}}$.

We find that in the regimes realised by the organic materials in regions I and II of the phase diagram of Fig.~2 of the main text, the contribution to $\phi_{\text{exc}}$ due to the envelope part $A_{\text{exc}}^{\text{en}}(Q)$ vanishes, whereas the single-particle contribution $A_{\text{exc}}^{\text{sp}}(Q)$ due to the band geometry of the constituent electrons and holes (see ``Riemannian geometry of excitons'' below), gives $\phi_{\text{exc}} = \pi$ on integrating. Notably, the envelope contribution vanishes, as under the combination of inversion ($\mathcal{P}$) and time-reversal ($\mathcal{T}$) symmetries present in the considered materials (see also ``Derivation of the bound on excitonic spread" below), the envelope function is a single-valued, \textit{real} smooth scalar function, which unlike the electron Bloch bands, cannot yield any non-trivial boundary terms due to $U(1)$-phase factors, on integrating over BZ. 

Moreover, beyond the numerical evaluation of the excitonic Berry phase and the dualisation argument detailed further below, we analytically model the topology of excitonic wavefunctions within the phase diagram by asserting an effective form of the excitonic envelope function. Explicitly, the exciton envelope function for the $t_2 > t_1$ regime, obtained by a qualitative fit to the Wannier equation solution, reads:
\beq{}
    \psi_{Q}(k) = \mathcal{N} \Big[ \cos (k+Q) e^{-(k+Q/2)^2/2} + \cos (k-Q) e^{-(k-Q/2)^2/2} \Big],
\eeq
with the normalisation constant $\mathcal{N}$.~With this expression, we obtain a vanishing envelope contribution ${\oint \dd Q~ A^{\text{en}}_{\text{exc}}(Q) = 0}$, confirming that the excitonic topology is determined by the topology of the underlying electron and hole states. Consistently, we show the smooth dependence of the excitonic envelope functions $\psi_{Q}(k)$ on the exciton momentum $Q$ for even and odd excitonic bands in Supplementary Fig.~\ref{fig:figS3}.

\begin{figure} [!t]
    \centering
    \includegraphics[width=\textwidth]{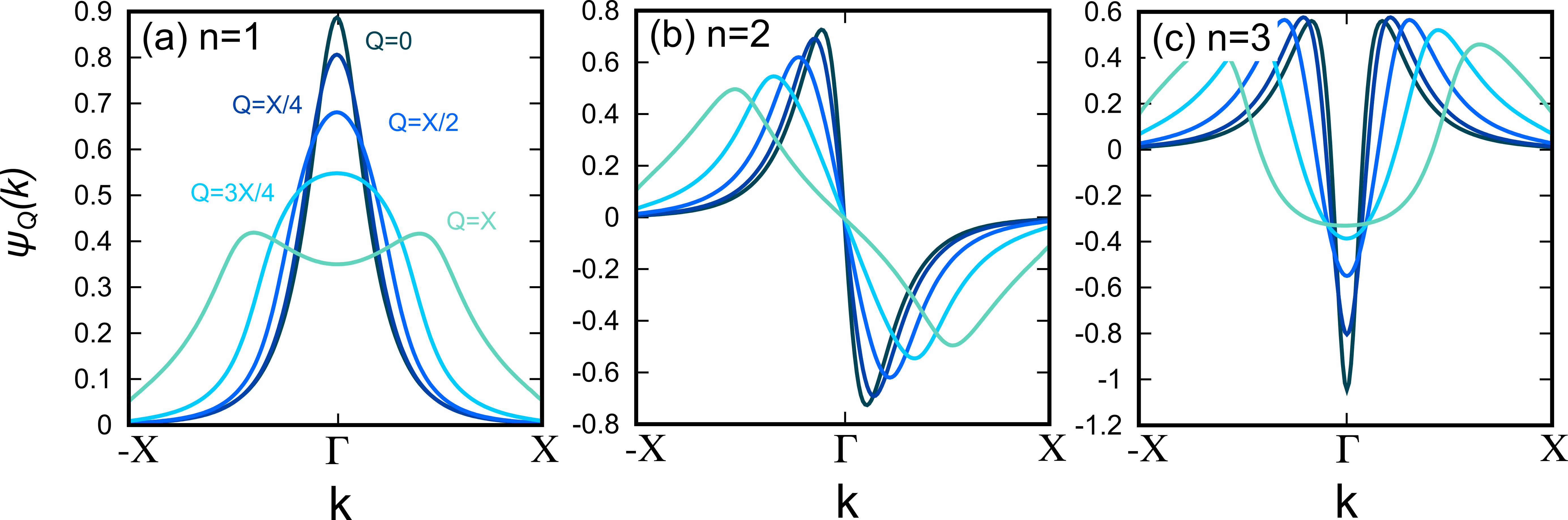}
    \caption{{\bf Excitonic envelope wavefunctions.} \textbf{(a)} Excitonic envelope function for the first excitonic band ($n=1$).  \textbf{(b)} Excitonic envelope function for the second excitonic band ($n=2$). \textbf{(c)} Excitonic envelope function for the third excitonic band ($n=3$). The excitonic envelope functions $\psi_Q (k)$ are shown for different exciton centre-of-mass momenta $Q$ within the momentum space range $(0, X)$. Here, $k =X$ is the momentum at the BZ edge and $k =\Gamma$ denotes the BZ centre. In the considered polyacenes, the excitonic envelope functions introduce no non-trivial excitonic Berry phase $\phi_\text{exc}$ contributions, with all the non-trivial topology of excitons being inherited from the topological electrons and holes.}
    \label{fig:figS3}
\end{figure}

\section*{Supplementary Note 3: Dualisation of the interacting Hamiltonian}

To reflect the physics of the (screened) density-density interactions, we truncate the long-range interaction expansion at the level of the nearest-neighbor interactions $U_1 \equiv U^{AB}_1$ and $U_2 \equiv U^{AB}_2$\,\cite{davenport2024interactioninduced}, assuming that $U_1, U_2 \gg U_{i>2}$. In the limit $t_1,t_2 \gg U_1, U_2$, we obtain the standard SSH model for electronic bands, and $\oint \dd Q\,A^{\text{en}}_{\text{exc}}(Q) = 0$. In this case, the entire contribution to excitonic topology can only emerge from the electrons and holes $\oint \dd Q~ A^{\text{sp}}_{\text{exc}}(Q) = \pi$. Intuitively, this corresponds to the formation of excitons on the same site where the constituent electrons and holes reside. 

In the opposite limit, $U_1,U_2 \gg t_1, t_2$, the Hamiltonian reduces to:
\beq{}
\begin{split}
    H = \frac{U_1}{2} \sum_{j} [ n_{A,j} n_{B,j} + n_{B,j} n_{A,j} ]  + \frac{U_2}{2} \sum_{j} [ n_{A,j+1} n_{B,j} + n_{B,j} n_{A,j+1} ].
\end{split}
\eeq
Notably, we recognise that $n_{B,j} = c^{\dagger}_{B,j} c_{B,j} = c^{\dagger}_{B,j} h^{\dagger}_{B,j} \equiv b^{\dagger}_{B,j}$, where $h^{\dagger}_{B,j}$ is a creation operator for a hole, and in combination with the creation operator of an electron $c^{\dagger}_{B,j}$, gives a localised exciton creation operator $b^{\dagger}_{B,j}$. Similarly, for sites $A$ we write: $n_{A,j} = c^{\dagger}_{A,j} c_{A,j} = h_{A,j} c_{A,j} \equiv b_{A,j}$.  On relabelling $\frac{U_{1/2}}{2} \rightarrow -t^{\text{exc}}_{1/2}$, we obtain a dualised SSH Hamiltonian for excitons:
\beq{}
\begin{split}
    H = -t^{\text{exc}}_{1} \sum_{j} b^{\dagger}_{B,j} b_{A,j} + \text{h.c.} -t^{\text{exc}}_{2} \sum_{j} b^{\dagger}_{{B+1},j} b_{A,j} + \text{h.c.},
\end{split}
\eeq
where the Hubbard interactions between the electrons and holes on different sites effectively drive exciton hopping, or, in other words, exciton delocalisation. Notably, as $t^{\text{exc}}_{1/2}$ are not independent from the electronic $t_{1/2}$, we observe no regions where topological excitons would be obtained in the lowest excitonic band from trivial electrons and holes in~\cite{davenport2024interactioninduced} in the phase diagram, which would correspond to an additional topological regime, cf. Fig.~3 in the main text. The reason for the lack of such independence in the real material context (unlike in an abstract tight-binding model construction) is that both $t_{1/2}$ and $U_{1/2}$ are given by the matrix elements of the same Wannier states and local Hamiltonians, as well as (Coulomb) interaction potentials, and are therefore not independent. It should be noted here that the interactions manifestly do $not$ break the inversion symmetry, by construction of the model.

Importantly, we stress that large dielectric screening $(\epsilon_s \gg 1)$ can suppress the interaction strengths $U_1, U_2$. In the studied experimentally relevant organic systems, the dielectric screening allows to effectively control, i.e. suppress, the interaction strengths $U_1, U_2$. In that context, $U_2$ might be significantly lower than $U_1$ ($U_1 \gg U_2$), as long as $t_1 > t_2$, which imposes that the charge density over the bonds with enhanced hoppings is larger, consistently with the Wannier equation, and with the chemistry of the material\,\cite{Clas5}. At the DFT level, the Hartree term driven by the local mean-field electric charge density requires that $t_2 > t_1$ results in $U_2 > U_1$. It should be stressed that, on the contrary, the exchange interaction in the considered organic materials is negligible, therefore, the Hartree term dominates.

\section*{Supplementary Note 4: Controlling excitonic topology with strain}

First principles results on the impact of strain on the structure and single-particle electronic band structure of the polyacenes are illustrated in Supplementary Fig.\,\ref{fig:figS4}. In Supplementary Fig.\,\ref{fig:figS4}(a), we show the change in the molecular polymer unit as a result of compressive (red) and tensile (green) strain. The impact of strain is primarily localised on the central carbon ring leading to a modulation of the $\text{C}$--$\text{C}$ (carbon-to-carbon) distance, both along and perpendicular to the polymer chain.

\begin{figure}[!t]
    \centering
    \includegraphics[width=\linewidth]{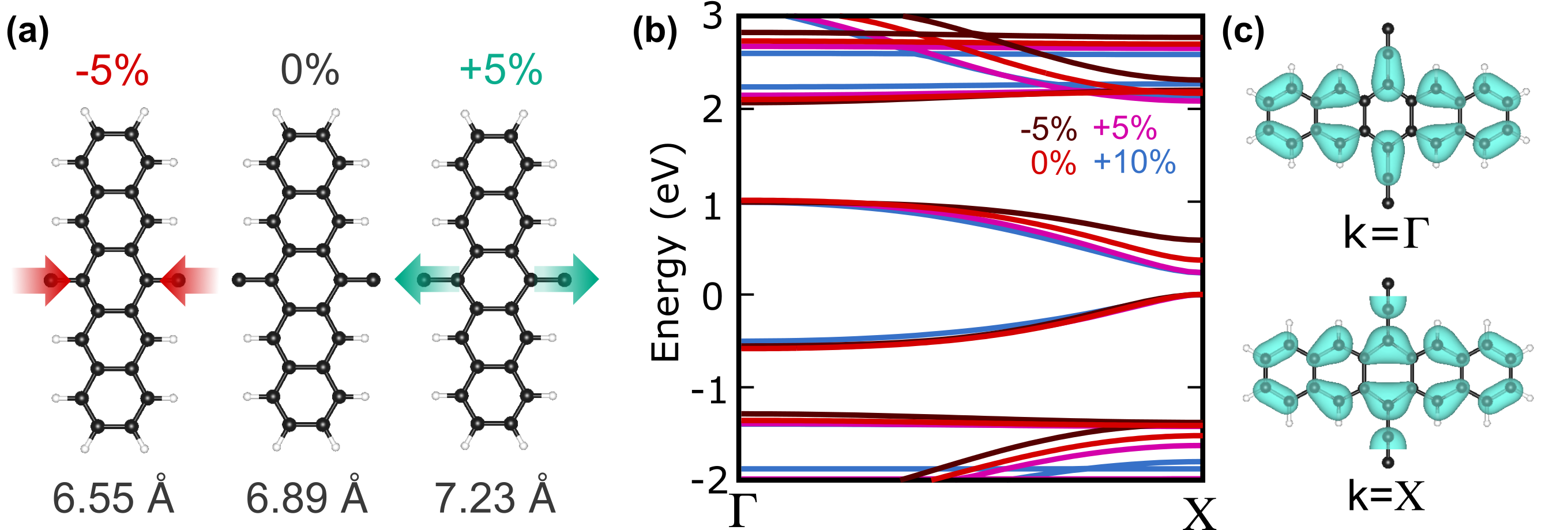}
   \caption{{\bf First principles calculations and the impact of strain on polyacenes.} \textbf{(a)} Impact of uniaxial strain $|\gamma| \leq 10\%$ applied along the polymer chain axis on the crystal structure of polypentacene. The percentage values of the strain, $\gamma$, and the values of the corresponding lattice parameters, $a'$, are provided above and below the strained molecular structures, respectively. \textbf{(b)} Electronic band structure of polypentacene at different strains: $\gamma = 0 \%$ (red), $\gamma = -5 \%$ (black), $\gamma = 5 \%$ (purple), $\gamma = 10 \%$ (blue) are shown. \textbf{(c)} Valence band electronic orbital at $k=\Gamma$ and $k=X$. The change in charge centre compared to the unstrained pentacene [c.f Fig.~6(e) of the main text] is characteristic of the polyanthracene [c.f Fig.~6(d) of the main text].}
   \label{fig:figS4}
\end{figure}

The electronic bandstructure of strained polypentacene is shown in Supplementary Fig.\,\ref{fig:figS4}(b), with both compressive and tensile strain. The VBM is set to $0$\,eV in all cases for comparison. For $5\%$ compressive strain the bandwidth is reduced and the band gap increases while for $5\%$ tensile strain the opposite is observed. This is in accord with the topological electrons and holes in polypentacene, as isolated molecules should always be topologically trivial, and large tensile strains should move the band structure towards that of a topologically trivial state. At $10\%$ tensile strain the band structure is quantitatively similar to that of $5\%$ strain. However, upon inspecting the electronic orbitals, Supplementary Fig.\,\ref{fig:figS4}(c), a stark change in the orbital shape is seen at $k=X$ as compared to the unstrained molecule [c.f., Fig.~6(e)]. Interestingly, the overall shape and structure of these orbitals in the central carbon ring is close to those observed in polyanthracene showing that the topological electronic states in highly strained polypentacene have become trivialised. This is realised as the trivial exciton shown in Fig.~2(b) of the main text ($t_1>t_2$). 

From an analytical perspective, we can model the effect of strain as renormalising the hopping parameters as,
\beq{}
    t_{1/2}(\gamma) = t_{1/2}(\gamma = 0) e^{-\gamma C_{1/2}}
\eeq
where $\gamma = \Delta a/a = (a'-a)/a$ is the uniaxial strain strength, captured by the change in the lattice parameter $\Delta a= a'-a$, that changes from the original value $a$ to $a'$. $C_{1/2}$ are phenomenological constants capturing distinct scaling of $t_{1/2}$ under the action of strain. Importantly, $C_{1} > C_{2}$ ($C_{1} < C_{2}$) is a necessary condition for an excitonic topological phase transition to happen, if $t_1 > t_2$ ($t_1 < t_2$). Otherwise, the crossover of hoppings, which induces the topological transition, cannot occur. Notably, a similar crossover associated with an electronic strain-induced topological phase transition was recently observed and characterised in graphene nanoribbons\,\cite{Tepliakov2023}.

\section*{Supplementary Note 5: Riemannian geometry of excitons}

In this section, we further elaborate on the Riemannian structure of the Bloch bundles of the excitonic wavefunctions and the associated excitonic quantum metric. We start by recognizing that we can equip a set, or bundle, of excitonic states $\ket{u^{\text{exc}}_\textbf{Q}}$ with a Hermitian metric, known otherwise as a quantum-geometric tensor (QGT). Explicitly, we write\,\cite{bouhon2023quantum}:
\beq{}
 Q^{\text{exc}}_{ij} = \bra{\partial_{Q_i} u^{\text{exc}}_\textbf{Q}} 1 - \hat{P} \ket{\partial_{Q_j} u^{\text{exc}}_\textbf{Q}},\eeq
where $\hat{P} = \ket{u^{\text{exc}}_\textbf{Q}}\bra{u^{\text{exc}}_\textbf{Q}}$ is a projector onto the excitonic band of interest. The excitonic QGT can be decomposed in terms of Riemannian (real) and symplectic (imaginary) parts. The imaginary part defines an excitonic Berry curvature, which can be non-trivial only in two- or higher-dimensional systems, whereas the real part defines an excitonic Riemannian metric $g^{\text{exc}}_{ij} \equiv \mathfrak{Re}~Q^{\text{exc}}_{ij}$, which can be non-vanishing even in one-dimensional systems.

The metric defined as the real part of the QGT explicitly reads:
\beq{}
    g^{\text{exc}}_{ij} = \frac{1}{2} \left[ \bra{\partial_{Q_i} u^{\text{exc}}_\textbf{Q}} \hat{Q} \ket{\partial_{Q_j} u^{\text{exc}}_\textbf{Q}} + \bra{\partial_{Q_j} u^{\text{exc}}_\textbf{Q}} \hat{Q} \ket{\partial_{Q_i} u^{\text{exc}}_\textbf{Q}} \right],
\eeq
where $\hat{Q}=1-\hat{P}$. In the one-dimensional case, which we consider in this work, there exists a single excitonic metric component of interest:
\beq{eq::gxx}
\begin{split}
    g^{\text{exc}}_{xx} = \frac{1}{2} \Big[ \bra{\partial_{Q_x} u^{\text{exc}}_{Q_x}} \ket{\partial_{Q_x} u^{\text{exc}}_{Q_x}} - \bra{\partial_{Q_x} u^{\text{exc}}_{Q_x}} \ket{u^{\text{exc}}_{Q_x}} \bra{ u^{\text{exc}}_{Q_x}} \ket{\partial_{Q_x} u^{\text{exc}}_{Q_x}} + \text{c.c.} \Big] \\= \bra{\partial_{Q_x} u^{\text{exc}}_{Q_x}} \ket{\partial_{Q_x} u^{\text{exc}}_{Q_x}} - \bra{\partial_{Q_x} u^{\text{exc}}_{Q_x}} \ket{u^{\text{exc}}_{Q_x}} \bra{ u^{\text{exc}}_{Q_x}} \ket{\partial_{Q_x} u^{\text{exc}}_{Q_x}}.
\end{split}
\eeq
We note that, on taking normalised eigenvectors $\bra{u^{\text{exc}}_\textbf{Q}} \ket{u^{\text{exc}}_\textbf{Q}} = 1$, and using the identity $0 = \partial_Q (\bra{u^{\text{exc}}_\textbf{Q}} \ket{u^{\text{exc}}_\textbf{Q}}) = \bra{u^{\text{exc}}_\textbf{Q}}\ket{\partial_Q u^{\text{exc}}_\textbf{Q}} + \bra{\partial_Q u^{\text{exc}}_\textbf{Q}}\ket{u^{\text{exc}}_\textbf{Q}}$, we obtain $\bra{u^{\text{exc}}_\textbf{Q}}\ket{\partial_Q u^{\text{exc}}_\textbf{Q}} = -\bra{\partial_Q u^{\text{exc}}_\textbf{Q}}\ket{u^{\text{exc}}_\textbf{Q}}$. Hence, in the one-dimensional case of interest, and taking $Q \equiv Q_x$, i.e. dropping $x$-index for simplicity, we can also write,
\beq{eq:gAA}
    g^{\text{exc}}_{xx} = \bra{\partial_{Q} u^{\text{exc}}_{Q}} \ket{\partial_{Q} u^{\text{exc}}_{Q}} -  A_{\text{exc}} A_{\text{exc}},
\eeq
where $A_{\text{exc}} = \ii \bra{u^{\text{exc}}_{Q}}\ket{\partial_{Q} u^{\text{exc}}_{Q}}$.

In general, the exciton quantum metric can be decomposed as:
\beq{}
    g^{\text{exc}}_{ij}(\vec{Q}) = g^{\text{en}}_{ij}(\vec{Q}) + g^{\text{sp-en}}_{ij}(\vec{Q}) + g^{\text{sp}}_{ij}(\vec{Q}),
\eeq
with the envelope contribution:
\beq{}
    g^{\text{en}}_{ij}(\vec{Q}) = \frac{1}{2} \sum_{\kv} [\partial_{Q_j} \psi^*_\vec{Q}(\kv) \partial_{Q_i} \psi_\vec{Q}(\kv) + \partial_{Q_i} \psi^*_\vec{Q}(\kv) \partial_{Q_j} \psi_\vec{Q}(\kv)]-A^{\text{en},i}_{\text{exc}} A^{\text{en},j}_{\text{exc}},
\eeq
the single-particle contribution:
\beq{}
    g_{ij}^{\text{sp}}(\vec{Q}) = \frac{1}{4} \sum_{\kv} |\psi_\vec{Q}(\kv)|^2 \Big[ g_{ij}^{\text{e}}(\kv + \vec{Q}/2) + g_{ij}^{\text{h}}(\kv - \vec{Q}/2) \Big]-A^{\text{sp},i}_{\text{exc}} A^{\text{sp},j}_{\text{exc}},
\eeq
and the mixed single-particle/envelope term:
\beq{}
    g_{ij}^{\text{sp-en}}(\vec{Q}) = \frac{1}{2} \sum_{\kv} \Big[ \partial_{Q_i} |\psi_\vec{Q}(\kv)|^2 \partial_{Q_j} A^{\text{sp},i}_{\text{exc}} + \partial_{Q_i} |\psi_\vec{Q}(\kv)|^2 \partial_{Q_i} A^{\text{sp},j}_{\text{exc}} \Big] - \Big[ A^{\text{sp},i}_{\text{exc}} A^{j}_{\text{exc}} + A^{i}_{\text{exc}} A^{\text{sp},j}_{\text{exc}} \Big].
\eeq
We note that the last term in each of the excitonic metric decomposition terms is  the corresponding contribution from the excitonic Berry connection term, i.e. the second term from the decomposition of Supplementary Eq.~\eqref{eq:gAA}. In the above, $g_{ij}^{\text{e/h}}(\kv)$ denote the individual single-particle quantum metrics of the electrons and holes. As mentioned in the main text, the envelope term $g^\text{en}_{ij}$ becomes significantly enhanced in the higher excitonic bands, as it involves more pronounced variations of the corresponding excitonic envelope functions, see Fig.~5 in the main text for reference.

As a concluding remark, we note that the Riemannian geometry of the excitons can be controlled with the dielectric screening, see Supplementary Fig.~\ref{fig:figS5}.

\begin{figure}[!h]
    \centering
    \includegraphics[width=0.5\linewidth]{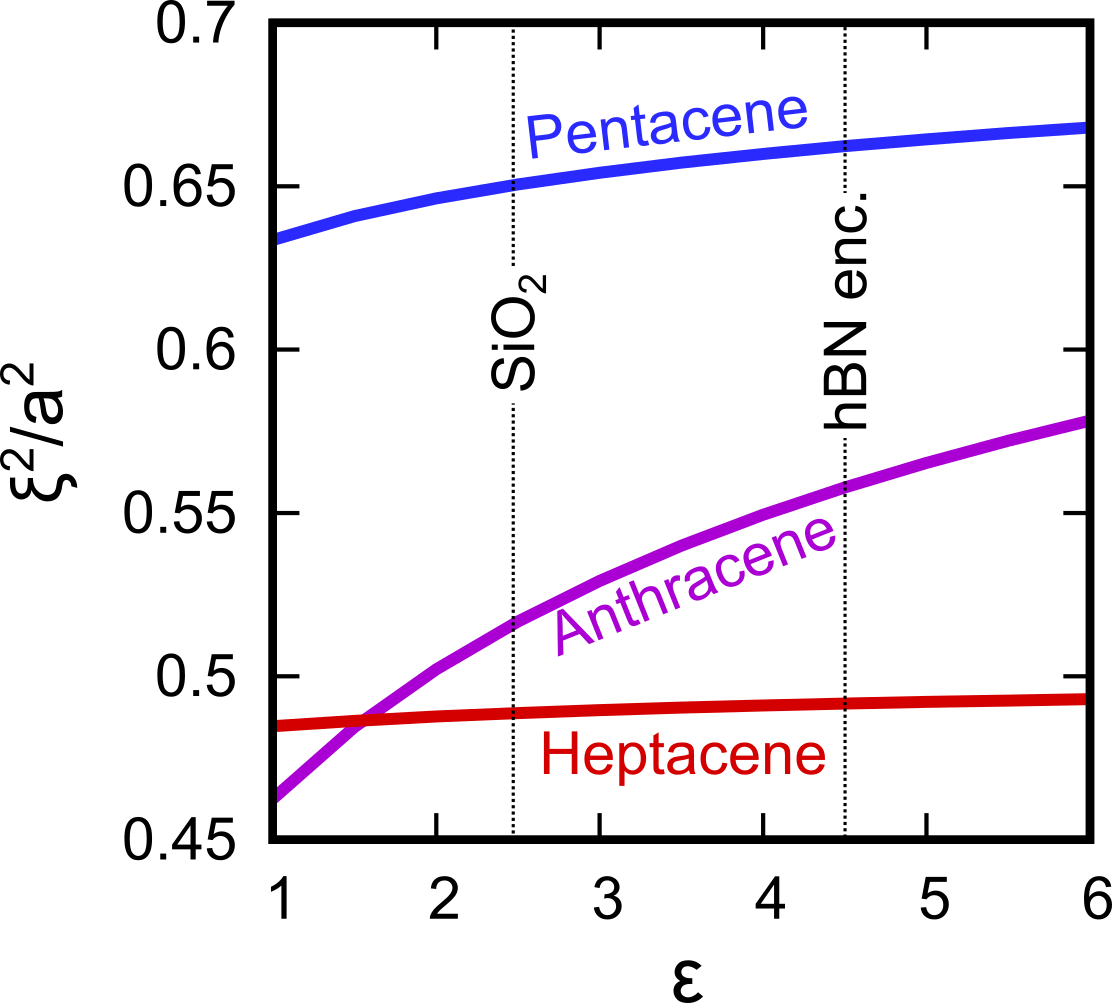}
   \caption{{\bf Controlling excitonic geometry with dielectric environment}. Impact of background dielectric screening on the excitonic variance for  polyanthracene (trivial, purple), polypentacene (topological, blue) and polyheptacene (red). Some typical dielectrics (SiO$_2$ substrate, hBN encapsulation) are marked with vertical lines.}
   \label{fig:figS5}
\end{figure}

While changing the dielectric screening does not result in  any excitonic topological phase transitions, because of $t_1$ being correlated with $U_1$, and $t_2$ being correlated with $U_2$ within the studied materials, the geometry of the excitons can be controlled, while satisfying the topological bound. This offers for controlling the associated exciton transport features~\cite{thompson2024topologicallyenhancedexcitontransport}, which were studied in detail in the subsequent work.

\section*{Supplementary Note 6: Derivation of the bound on excitonic spread}

The excitonic Riemannian metric encodes information about the localisation of the excitons, corresponding to the second moment, and in a one-dimensional context it is given by:
\beq{eq:metricXi}
    \xi^2 \equiv \text{Var} R = \langle R^2 \rangle - \langle R \rangle^2 = \frac{a}{2 \pi} \int_{\text{BZ}} \dd Q~ g^{\text{exc}}_{xx}(Q),
\eeq
where $\langle (\ldots) \rangle = \bra{w_0^{\text{exc}}} (\ldots) \ket{w_0^{\text{exc}}}$ and $\ket{w_{j}^{\text{exc}}} = \frac{a}{2\pi} \int_{\text{BZ}} dQ\,e^{- \ii Q ja} \ket{\psi^{\text{exc}}_Q}$ is an excitonic Wannier state with lattice vector $\vec{x} = ja$ labeling the unit cell of interest. Supplementary Eq.\,~\eqref{eq:metricXi} follows by recognizing that Supplementary Eq.\,\eqref{eq::gxx} implies:
\beq{}
\begin{split}
    g^{\text{exc}}_{xx} &= \bra{\partial_{Q} u^{\text{exc}}_{Q}} \ket{\partial_{Q} u^{\text{exc}}_{Q}} - \bra{\partial_{Q} u^{\text{exc}}_{Q}} \ket{u^{\text{exc}}_{Q}} \bra{ u^{\text{exc}}_{Q}} \ket{\partial_{Q} u^{\text{exc}}_{Q}} \nonumber \\ 
    &= \sum_{m} \bra{\partial_{Q} u^{\text{exc}}_{Q}} \ket{u^{\text{exc}}_{m Q}} \bra{u^{\text{exc}}_{m Q}} \ket{\partial_{Q} u^{\text{exc}}_{Q}} \nonumber \\
    &= \sum_m \bra{\psi^{\text{exc}}_{Q}} R \ket{\psi^{\text{exc}}_{m Q}} \bra{\psi^{\text{exc}}_{m Q}} R \ket{\psi^{\text{exc}}_{Q}} \nonumber \\
    &=  \bra{\psi^{\text{exc}}_{Q}} R ( 1 -\ket{\psi^{\text{exc}}_{ Q}} \bra{\psi^{\text{exc}}_{Q}} ) R \ket{\psi^{\text{exc}}_{Q}}.
\end{split}
\eeq
This result follows by recognizing that for $n \neq m$, we have $\bra{\psi^{\text{exc}}_{n Q}} R \ket{\psi^{\text{exc}}_{m Q}} = \ii \bra{u^{\text{exc}}_{n Q}} \ket{\partial_{Q} u^{\text{exc}}_{m Q}}$, with ${\ket{\psi^{\text{exc}}_{m Q}} = e^{\ii Q R} \ket{u^{\text{exc}}_{m Q}}}$, and where $m$ runs over all excitonic band indices apart from the index of the band $\ket{u^{\text{exc}}_{Q}}$ of interest. Moreover, we used the resolution of the identity ${1 = \sum_m \ket{\psi^{\text{exc}}_{m Q}} \bra{\psi^{\text{exc}}_{m Q}} + \ket{\psi^{\text{exc}}_{Q}} \bra{\psi^{\text{exc}}_{Q}}}$.
%
%
On changing the basis according to $\ket{w_0^{\text{exc}}} = \frac{a}{2\pi} \int_\text{BZ} \dd Q~ e^{i QR} \ket{u^{\text{exc}}_{Q}}$, and evaluating the matrix elements with the position operator $R$, we obtain the following relationship between the excitonic quantum metric $g^{\text{exc}}_{xx}$ and spread $\xi^2$:
\beq{}
    \frac{a}{2\pi} \int_\text{BZ} \dd Q~ g^{\text{exc}}_{xx} = \bra{w_0^{\text{exc}}} R^2 \ket{w_0^{\text{exc}}} - \bra{w_0^{\text{exc}}} R \ket{w_0^{\text{exc}}}^2 = \langle R^2 \rangle - \langle R \rangle^2 = \langle (R - \langle R \rangle)^2 \rangle = \text{Var} R \equiv \xi^2.
\eeq

We now derive the main bound on the excitonic spread, captured by the excitonic metric due to the excitonic invariant $P_{\text{exc}}$. The excitonic $\mathbb{Z}_2$ invariant under inversion ($\mathcal{P}$) symmetry, in the gauge admitting only $\phi_\text{exc} = 0/\pi$ under the symmetry, can be written as:
\beq{}
    P_{\text{exc}} = \frac{1}{\pi} \int^{\pi/a}_0 \dd Q~[ A_{\text{exc}}(Q) + A_{\text{exc}}(-Q)] = \frac{1}{\pi} \int^{2\pi/a}_0 \dd Q~ A_{\text{exc}}(Q).  
\eeq
Moreover, the invariant can be alternatively deduced from the high-symmetry points (HSPs) $Q=0$ and $Q=\pi$, analogously to how the topological invariants can be deduced from time-reversal invariant momenta (TRIMs) in the electronic topological insulators~\cite{Fu2007}. Equipped with the parities of the excitonic band at the HSPs, $\delta_i$, the excitonic invariant satisfies,
\beq{}
    (-1)^{P_{\text{exc}}} = \Pi_i \delta_i.
\eeq
Hence, $P_{\text{exc}} = 1$ manifestly requires different band parity eigenvalues at $Q =0$ and $Q = \pi$. We note, that in the studied organic materials, the excitonic states also satisfy the bosonic spinless time-reversal symmetry ($\mathcal{T}^2 = 1$), culminating in the spinless spatiotemporal inversion symmetry, $(\mathcal{PT})^2 = 1$~\cite{ bouhon2020geometric}. Therefore, topologically, the $P_{\text{exc}}$ invariant corresponds to the first Stiefel-Whitney characteristic class $w_1 \in \mathbb{Z}_2$~\cite{Naka, Ahn_2019SW, bouhon2020geometric} of the (real) excitonic Bloch bundle, as we also discuss later. We now utilise the topological invariant associated with the mentioned characteristic class to demonstrate the quantum-geometric bound.

Correspondingly, we use a Cauchy-Schwarz inequality $\Big| \int \dd Q~ f_1(Q) f_2(Q) \Big|^2 \leq \Big( \int \dd Q~ |f_1(Q)|^2 \Big) \Big( \int \dd Q~ |f_2(Q)|^2 \Big)$, with functions $f_1(Q) = 1$ and $f_2(Q) = A_{\text{exc}}(Q)$,
\beq{eq:ineq}
     P^2_{\text{exc}} = \frac{1}{\pi^2} \Bigg|\int^{2\pi/a}_0 \dd Q ~A_{\text{exc}}(Q)\Bigg|^2 \leq \frac{1}{\pi^2} \Bigg( \int^{2\pi/a}_0 \dd Q' \Bigg) \Bigg( \int^{2\pi/a}_0 \dd Q ~\Big|A_{\text{exc}}(Q) A_{\text{exc}}(Q)\Big| \Bigg) \leq \frac{2}{\pi a} \int^{2\pi/a}_0 \dd Q~ g^{\text{exc}}_{xx}(Q), 
\eeq
where $a$ is the lattice parameter. Here, we additionally recognise that in the maximally-smooth gauge extremising the overlaps of Bloch states between the neighbouring $Q$-points, $\bra{u^\text{exc}_Q}\ket{u^\text{exc}_{Q + \Delta Q}} \approx 1 + \bra{u^\text{exc}_Q} \ket{\partial_Q u^\text{exc}_Q} \Delta Q \rightarrow 1$, that obtains the maximally-localised Wannier functions~\cite{marzari2012maximally}: $\Big|A_{\text{exc}}(Q) A_{\text{exc}}(Q)\Big| = \bra{\partial_{Q} u^{\text{exc}}_{Q}} \hat{P} \ket{ \partial_{Q} u^{\text{exc}}_{Q}} \leq \bra{\partial_{Q} u^{\text{exc}}_{Q}} \hat{Q} \ket{\partial_{Q} u^{\text{exc}}_{Q}} = g^{\text{exc}}_{xx}(Q)$, i.e. the overlap of $\ket{u^{\text{exc}}_{Q}}$ and $\ket{\partial_{Q} u^{\text{exc}}_{Q}}$ is locally minimised, and is smaller than the overlap of $\ket{\partial_{Q} u^{\text{exc}}_{Q}}$ with all the other excitonic bands $\ket{u^{\text{exc}}_{m Q}}$ combined, as we also retrieve numerically. Intuitively, it should be noted that in the limit of a~real gauge admitted under $\mathcal{PT}$ symmetry, the former overlap approaches vanishingly small values locally, consistently with the metric evolving as $g^{\text{exc}}_{xx}(Q) = \bra{\partial_{Q} u^{\text{exc}}_{Q}} \hat{Q} \ket{\partial_{Q} u^{\text{exc}}_{Q}} \rightarrow \bra{\partial_{Q} u^{\text{exc}}_{Q}} \ket{\partial_{Q} u^{\text{exc}}_{Q}}$. Finally, on rearranging the derived inequality Supplementary Eq.~\eqref{eq:ineq}, 
we obtain,
\beq{}
    \xi^2 \equiv \frac{a}{2\pi} \int^{2\pi/a}_0 \dd Q~ g^{\text{exc}}_{xx}(Q) \geq \frac{a^2 P_{\text{exc}}^2}{4},
\eeq
showing that the maximally-localised excitonic Wannier functions support standard deviation associated with the spread: $\xi \geq a/2$, in the non-trivial phase with $P_{\text{exc}} = 1$. Mathematically, the non-trivial topological invariant $P_{\text{exc}} = 1$ associated with the non-trivial first Stiefel-Whitney class $w_1 \in \mathbb{Z}_2$ captures the orientability of a line subbundle of a real Bloch bundle~\cite{Panati2007}, and is a global property of the excitonic Bloch bundles considered in this work. Intuitively, the non-trivial topology captured by the characteristic class implies the non-trivial quantum geometry locally, as a result of the associated non-orientability condition~\cite{Naka}. 

In the absence of non-trivial $P_{\text{exc}} \in \mathbb{Z}_2$, as well as of any other topological invariants in a considered phase, i.e. in the presence of only trivial excitons, any possible enhancement of the excitonic equivalent to the minimal spread of trivial states ($\xi^2$) is expected to be unlikely, as $\xi^2$ is unconstrained by any lower bound. This assertion, more formally yet intuitively, follows from the definition of the quantum metric:
\beq{}
    g^\text{exc}_{xx}(Q) = \bra{\partial_{Q}u^\text{exc}_Q} (1- \ket{u^\text{exc}_Q} \bra{u^\text{exc}_Q}) \ket{\partial_{Q}u^\text{exc}_Q} = \bra{\partial_{Q}u^\text{exc}_Q} \hat{Q} \ket{\partial_{Q}u^\text{exc}_Q} = \bra{\partial_{Q}u^\text{exc}_Q} \hat{Q}^2 \ket{\partial_{Q}u^\text{exc}_Q} = \Big|\Big|\hat{Q} \ket{\partial_{Q}u^\text{exc}_Q} \Big|\Big|^2,
\eeq
where we employed a projector $\hat{Q} = 1- \ket{u^\text{exc}_Q} \bra{u^{\text{exc}}_Q}$ and used its idempotency, $\hat{Q} = \hat{Q}^2$, and Hermiticity $\hat{Q} = \hat{Q}^\dagger$ relations on connecting to the final norm $||...||$. In the case of the trivial topologies, the Bloch states $\ket{u^\text{exc}_Q}$ do not wind, hence $\ket{\partial_{Q}u^{\text{exc}}_Q} \rightarrow \textbf{0}$, with $\textbf{0}$, the norm-zero vector. Hence, $g^{\text{exc}}_{xx}  = ||\hat{Q} \ket{\partial_{Q}u^{\text{exc}}_Q} ||^2 \rightarrow ||\hat{Q} \textbf{0} ||^2 \rightarrow 0$, as the Frobenius norm of $\hat{Q}$ is smaller than for the identity matrix $1$
that sums over the projectors onto all bands, rather than onto a smaller number of bands, as  $\hat{Q} = 1- \hat{P}$, which on having subtracted $\hat{P} =\ket{u^{\text{exc}}_Q} \bra{u^{\text{exc}}_Q}$, excludes the projection onto the original excitonic band $\ket{u^{\text{exc}}_Q}$.

\section*{Supplementary Note 7: Numerical excitonic localisation}

Finally, in the light of the analytically derived quantum-geometric bound imposed by the non-trivial excitonic topology, we describe how the excitonic localisation§ that reflects the excitonic quantum geometry, can be retrieved numerically in a real material.

Here, we stress that to reflect the excitonic quantum geometry, the maximally-localised excitonic Wannier functions (MLXWFs)~\cite{PhysRevB.108.125118} need to be obtained, as the excitonic quantum metric reflects the \textit{maximal} localisation as a gauge-invariant quantity. In an arbitrary gauge, the excitonic Wannier functions can take a less-localised form than fixed by the gauge-invariant quantum metric $g^{\text{exc}}_{xx}(Q)$. Contrary to the gauge-independent/gauge-invariant excitonic quantum metric and maximally-localised basis, an arbitrary excitonic Wannier function basis does not bear any physical meaning, as a gauge-dependent object. Such scenario is in a one-to-one correspondence analogy to the case of the electronic Wannier functions~\cite{vanderbilt2018berry}.

As a consequence, in the context of a real material, it is essential to numerically obtain the gauge-independent excitonic maximal localisation~\cite{PhysRevB.108.125118}. Following the approach of Ref.~\cite{PhysRevB.108.125118}, this can be obtained by minimisation of a~localisation functional $F_{\text{exc}}$ that depends on the excitonic Wannier functions, $\ket{w^{\text{exc}}_0}$, 
\beq{}
    F_{\text{exc}} = \int^{\infty}_{-\infty} \dd R \bra{w^{\text{exc}}_0} R^2 \ket{w^{\text{exc}}_0} - \int^{\infty}_{-\infty} \dd R \Big[ \bra{w^{\text{exc}}_0} R \ket{w^{\text{exc}}_0} \Big]^2.
\eeq
For minimising the functional $F_{\text{exc}}$, the algorithm of {\sc Wannier90}~\cite{MOSTOFI20142309} can be directly utilised~\cite{PhysRevB.108.125118}, on paralleling the standard method for minimising the spread of the electronic Wannier functions~\cite{PhysRevB.56.12847, marzari2012maximally, vanderbilt2018berry}.

Consistently with the localisation procedure described in Ref.~\cite{marzari2012maximally}, the maximally-localised basis that minimises the localisation functional $F_{\text{exc}}$, can be numerically retrieved from sampling the possible $U(1)$ gauges $[U(Q) = e^{i \alpha(Q)}]$ of the isolated individual Fourier-transformed excitonic Bloch bands $\ket{u^{\text{exc}}_Q}$,
\beq{}
  \ket{w^{\text{exc}}_j} = \frac{a}{2\pi} \intBZ \dd Q~ e^{-i Q (R)_j} U(Q) \ket{u^{\text{exc}}_Q}.
\eeq
In the context of polyacenes, we further observe that such basis corresponds to the smoothest realisable waveform $w^{\text{exc}}_0 (R) = \bra{R}\ket{w^{\text{exc}}_{0}}$, as quantified by the averaged analytic curvature of the excitonic Wannier function $\langle \partial^2_x w^{\text{exc}}_0 (R) \rangle$.

\section*{Supplementary References}
\bibliographystyle{naturemag}
\bibliography{references}